\documentclass[12pt,letterpaper]{article}

\usepackage{datetime}
\usepackage{amsmath,amssymb,array,calc,rotating,epsfig,psfrag, amscd}
\usepackage{color}
\usepackage[linktoc=page,
pdfstartview=FitV,
    bookmarksopen=true]{hyperref}
\usepackage[left=2cm,top=1cm,right=3cm,nohead]{geometry}

\newcommand{\bea}{\begin{eqnarray}}
\newcommand{\eea}{\end{eqnarray}}
\newcommand{\be}{\begin{equation}}
\newcommand{\ee}{\end{equation}}
\newcommand{\barr}{\begin{array}}
\newcommand{\earr}{\end{array}}
\newcommand{\del}{\partial}
\newcommand{\tphi}{\tilde{\phi}}
\newcommand{\txi}{\tilde{\xi}}

\newcommand{\csch}{{\rm csch \,}}
\newcommand{\sech}{{\rm sech \,}}
\newcommand{\non}{\nonumber}
\definecolor{cardinal}{rgb}{0.6,0,0}
\definecolor{darkgreen}{rgb}{0,0.5,0}
\definecolor{golden}{rgb}{0.92, 0.7, 0}
\definecolor{midnight}{rgb}{0, 0, 0.5}
\definecolor{darkblue}{rgb}{0.2, 0, 0.8}

\newcommand{\beq}{\begin{equation}\begin{aligned}}
\newcommand{\eeq}{\end{aligned}\end{equation}}
\newcommand{\nn}{\nonumber}

\numberwithin{equation}{section}

\begin{document}

\thispagestyle{empty}
\begin{flushright}
 IPhT-t11/199
 \end{flushright}

\vspace{0.5cm}
\begin{center}
\baselineskip=13pt {\LARGE \bf{Metastable Vacua and the \\Backreacted
    Stenzel Geometry\\}}
 \vskip1.5cm 
\renewcommand{\thefootnote}{\fnsymbol{footnote}}
Stefano Massai\footnote{e--mail: \href{mailto:stefano.massai@cea.fr}{\texttt{stefano.massai@cea.fr}}}  \\
\vskip0.5cm
\textit{Institut de Physique Th\'eorique,\\
CEA Saclay, CNRS URA 2306,\\
F-91191 Gif-sur-Yvette, France}\\
\vskip0.5cm
\end{center}
\vskip1.5cm
\begin{abstract}
We construct an M--theory background dual to the metastable
state recently discussed by Klebanov and Pufu, which corresponds to
placing a stack of anti--M2 branes at the tip of a warped Stenzel space. With this purpose we analytically solve for the linearized
non--supersymmetric deformations around the warped Stenzel space, preserving the $SO(5)$ symmetries of the supersymmetric background, and
which interpolate between the IR and UV region.
We identify the supergravity
solution which corresponds to a stack of $\bar{N}$ backreacting anti--M2 branes
by fixing all the 12 integration constants in terms of $\bar{N}$.
While in the UV this solution has the desired features to
describe  the conjectured
metastable state of the dual (2+1)--dimensional theory, in the IR it
suffers from a singularity in the four--form flux, which we describe in some details.
\end{abstract}

\newpage

\tableofcontents

%%%%%%%%%%%%%%%%%%%%%%%%%%%%%%%%%%%%%%%%%%%%%
 \section{Introduction}
%%%%%%%%%%%%%%%%%%%%%%%%%%%%%%%%%%%%%%%%%%%%%
\setcounter{page}{1}
\renewcommand{\thefootnote}{\arabic{footnote}}\setcounter{footnote}{0}

The work of Intriligator, Seiberg and Shih~\cite{Intriligator:2006dd}
has drawn attention to mechanisms of metastable supersymmetry
breaking in quantum field theories. Since the constructions of such
states involve strongly coupled regimes, it is natural to address the
study of this phenomenon in stringy realizations of the supersymmetric
theories and indeed there exist many corners of the string theory web where
these constructions have been proposed. 
In Type IIA string theory one can try to build D-brane models in order
to reproduce metastability in the field theories engineered on the
brane world volumes; however, it turns out that the probe brane
picture (i.e. $g_s =0$) is too naive and once the backreaction of the
branes is taken into account these systems can fail to reproduce
metastable states of the supersymmetric theories~\cite{Bena:2006rg}. 

Another approach is to consider brane realisations which extend the
AdS/CFT correspondance to non--conformal or less--supersymmetric
theories and try to construct metastable states in this context. One
way to achieve  this is to start by a configuration of branes placed
at some Calabi-Yau singularity and consider the supergravity
solution obtained after smoothing the singular point. The most well
known example in Type IIB string theory is the
Klebanov--Strassler (KS) background~\cite{Klebanov:2000hb}. In this setting the first
evidence of a metastable state in the $SU(M(k+1)-p)\times SU(Mk-p)$ theory was given in~\cite{Kachru:2002gs,
  DeWolfe:2004qx}, with a construction that involves placing a stack
of anti--D3 branes in the KS geometry. These branes are attracted to
the tip of the deformed conifold and polarize into an NS5--brane due to the Myers
effect~\cite{Myers:1999ps}. In a probe analysis, it was shown~\cite{Kachru:2002gs}
that this state is metastable and long-lived if the number $p$ of anti--D3
branes is small ($\sim$ 8\%) compared to the $M$ units of R-R 
3-form flux of the unperturbed supersymmetric background. It is an important question whether this picture remains
valid once the backreaction of the anti--D3 branes on the KS geometry
is taken into account.
Constructing the full backreacted supergravity solution is a difficult
task, but one can make progress if some simplification is adopted,
namely i) smearing the sources on the $S^3$ of the deformed
conifold and ii) working in perturbation theory around
the supersymmetric background, at first order in the parameter
$p/M$. This simplified (but still difficult enough) problem was
investigated in~\cite{Bena:2009xk,Bena:2011wh}, where the full
linearized backreaction of the anti--D3 branes on the KS geometry has
been constructed\footnote{This problem was also adressed
  in~\cite{Dymarsky:2011pm}.}. 

It is of obvious interest to address the same question in different
contexts where metastable states are conjectured to exist, by string theory arguments, in supergravity backgrounds dual to strongly
coupled field theories.

In this paper we study the case of an $AdS_4/CFT_3$ correspondence,
which involves an $\mathcal{N}=2$ supersymmetric (2+1)--dimensional theory, whose
supergravity dual is $AdS_4 \times V_{5,2}$, where $V_{5,2}$ is the
7--dimensional Sasaki--Einstein space $V_{5,2}=SO(5)/SO(3)$. Recently, a gravity dual for a long-lived metastable state has been proposed in~\cite{Klebanov:2010qs}
based on the probe analysis, by placing a stack of anti--M2 branes at the
tip of the warped M--theory background with transverse Stenzel
space~\cite{Stenzel:1993}. Here, the analogue of the KS solution is
the supersymmetric solution of Cveti\v{c}, Gibbons, L\"u and
Pope (CGLP)~\cite{Cvetic:2000db}; indeed, the 8--dimensional Stenzel
space is a part of a family of Ricci--flat solutions parametrized by
the dimension $n$, which include the deformed conifold for $n=6$. The
mechanism for which the false vacuum decays is similar to
the KPV process~\cite{Kachru:2002gs}: the anti--M2 branes fall in the
warped throat and at the tip they polarize into M5--branes wrapping an
$S^3 \subset S^4$. The probe analysis of~\cite{Klebanov:2010qs} shows
that this state is metastable if $p/\tilde{M} \leq 0.054$, where
$\tilde{M}$ is the number of units of the 4--form flux of the CGLP background.

The effects of the backreaction of the
anti--M2 branes on the transverse geometry have been studied in~\cite{Bena:2010gs}, where the linearized equations that
govern the first--order backreacted solution have been solved
implicitly in terms of integrals by using the first--order formalism introduced by
Borokhov and Gubser~\cite{Borokhov:2002fm} and the full solution was
presented separately in the small and large radius limit. The main purpose of
that work was to study the IR behavior of the perturbed solution, and the conclusion of
this analysis was similar to the anti--D3 case, namely that the
conjectured solution dual to the metastable state exhibits certain
singularities which in the anti--M2 case lead to a divergent action in the
IR\footnote{See~\cite{Giecold:2011gw} for a similar analysis in a Type
IIA context.}. To decide whether this
singularity is admissible or not is a difficult task, and the answer
is clearly beyond the linearized approximation. One way to proceed is
to connect the IR and the UV region and to see if the conjectured
solution eventually develops problems in the ultraviolet. In this
perspective, it is clearly interesting to perform such an analysis in the
anti--M2 brane configuration, which in the IR can be thought as the
M-theory generalization of the Type IIB KS solution, but has a rather
different behavior in the UV. For example, in the M-theory background there
is not a logarithmic running of the charge, which is an important
feature of the KS background, and was crucial in the analysis of the
backreaction performed in~\cite{Bena:2011wh}.
In this paper we perform the analysis outlined above and by extending the results
of~\cite{Bena:2010gs} we present the full analytic solution
of the linearized supergravity equations which describe the most general
non--supersymmetric deformation of the warped Stenzel space compatible
with the symmetries of the CGLP background. With this result we are
able to study the effects of IR boundary conditions on the ultraviolet
behavior of the supergravity modes and we identify the unique solution
which has the desired features to describe anti--M2 branes in the CGLP
background  (leaving open the issue of the singularity discussed
above).\\

This paper is organized as follows. In Section 2 we review the computational
formalism and we solve analytically the system of
first--order differential equations governing
perturbations around the CGLP supersymmetric solution. Our full solution,
which is shown in Appendix~\ref{appelliptic},  
contains few single integrals that cannot be explicitly performed,
but they can easily be handled with numerical integration. In Section 3 we
show the expansions of our solution in the IR and in the UV region in
terms of a set of twelve integration constants denoted $(X_a, Y_a)$. In
Section~\ref{secchargeM2brane} we discuss the various charges in the
Stenzel background and we identify the perturbation due to the
presence of M2 branes. In Section~\ref{M2bc}
we impose the boundary conditions that arise from placing a stack of
anti--M2 branes at the tip ($r=0$) of the geometry and we discuss the
problems associated to an infrared singularity in the fluxes. We then summarize the
asymptotic behavior of the anti--M2 solution, which is expressed in
terms of the number of anti--M2 branes. As a check
of our boundary conditions, we compute the force exerted on a probe M2
brane and we show that it agrees with the one derived from the
brane/antibrane potential (which we review in Appendix~\ref{appforce}). We end with
a discussion in Section 6.

%%%%%%%%%%%%%%%%%%%%%%%%%%%%%%%%%%
\section{Linearized equations and their solutions}
%%%%%%%%%%%%%%%%%%%%%%%%%%%%%%%%%%

The linearized equations governing the deformations around the warped
Stenzel space have been derived in~\cite{Bena:2010gs} by using the
Borokhov--Gubser~\cite{Borokhov:2002fm} first--order formalism. 
We use the ansatz for the $SO(5)$--invariant supergravity solution
of~\cite{Klebanov:2010qs}\footnote{A more general solution which
  includes this $SO(5)$--invariant ansatz has been constructed in~\cite{Giecold:toappear} .}:
\begin{equation}\label{metric}
ds^2_{11} = e^{-2z(r)} dx_{\mu}dx^{\mu} + e^{z(r)}\big[ e^{2\gamma (r) }
dr^{2} + e^{2\alpha (r)} \sigma_i^2 + e^{2\beta (r)}
\tilde{\sigma}_i^2 + e^{2 \gamma (r) } \nu^2 \big]\ ,
\end{equation}
where $\sigma_i$, $\tilde{\sigma}_i$ ($i=1,2, 3$) and $\nu$ are the 1--forms in the
coset $SO(5)/SO(3)$ and $\mu = 0,1,2$. The four--form field strength $G_4$ is given by
\begin{equation}\label{fluxes}
 G_4= d K(r) \wedge dx^0 \wedge dx^1 \wedge dx^2 + m\, F_4 \, , 
 \end{equation}
where the internal flux $F_4$ is parametrized by 
\begin{align}
F_4 &= f'(r) dr\wedge\tilde{\sigma}_1\wedge\tilde{\sigma}_2 \wedge
\tilde{\sigma}_3+ h'(r)\, \epsilon^{ijk}dr\wedge\sigma_i\wedge \sigma_j
\wedge \tilde{\sigma}_k \\
&\quad+\frac12 (4h(r)-f(r))\epsilon^{ijk}\nu\wedge \sigma_i \wedge
\tilde{\sigma}_j \wedge\tilde{\sigma}_k
 -6\,h(r)\,\nu\wedge\sigma_1\wedge\sigma_2\wedge\sigma_3 \, ,  \non 
\end{align}
and the function $K(r)$ is fixed in terms of the other functions by
the equation of motion
\begin{equation}\label{eomG4}
d\star G_4 = \frac12 G_4 \wedge G_4\, .
\end{equation}
The method introduced in~\cite{Borokhov:2002fm} relies on the
existence of a superpotential $W$ defined such that its square gives the potential, namely
 \begin{equation}
 V(\phi) = \frac{1}{8}\, G^{ab}\, \frac{\del W}{ \del \phi^a}\, \frac{\del W}{ \del \phi^b} \, .
 \end{equation}
 We consider an expansion for the fields $\phi^a$ ($a=1,...,n$) around
 the supersymmetric background
\begin{equation}
 \label{split}
 \phi^a = \phi^a_0 + \phi^a_1(X) + {\cal O}(X^2)\, ,
 \end{equation}
where $X$ represents the set of perturbation parameters, $\phi^a_1$ is
linear in them, and $\phi^a_0$ are the functions in the CGLP solution,
written explicitly in~\eqref{CGLPbackground}. We will denote the set of functions $\phi^a$, $a=1,...,6$ in the following order
\begin{equation} \label{phidef}
 \phi^a=(\alpha ,\beta ,\gamma ,z ,f,h) \, .
 \end{equation}
The first order formalism gives a set of $2n$ linear first--order
differential equations for the perturbations $\phi^a_1$ and their
conjugates $\xi^a$:
\begin{align}
\frac{d\xi_a}{d\tau} + \xi_b\, M^b{}_a(\phi_0) &= 0 \, ,  \label{xieq} \\
\frac{d\phi_1^a}{d\tau} - M^a{}_b(\phi_0)\, \phi_1^b &= G^{ab}\, \xi_b \label{phieq} \, ,
\end{align}
where
\begin{equation}
\label{xidef}
\xi_a \equiv G_{ab}(\phi_0)\, \left( \frac{d\phi_1^b}{d\tau} - M^b_{\ d}(\phi_0)\, \phi_1^d \right) \ , \qquad M^b{}_d\equiv\frac12 \, \frac{\partial}{\partial \phi^d}\, \left( G^{bc}\, \frac{\partial W}{\partial \phi^c} \right) \, .
\end{equation}
The equations~\eqref{phieq} are the definitions of the $\xi_a$,while
the $n$ equations ~\eqref{xieq} form a closed set and imply the
equations of motion~\cite{Borokhov:2002fm}. The functions $\xi_{a}$ should additionally satisfy the zero--energy condition
\begin{equation} \label{ZEC}
\xi_a\, \frac{d\phi_0^a}{d \tau} = 0 \, .
\end{equation}
The field--space metric in~\eqref{xidef} is 
\begin{align}
\label{fieldmetric}
G_{ab}\, \phi^{\prime a}\, \phi^{\prime b} = \frac12 e^{-\alpha -
  3(\beta+ z)}&\Big[ 3e^{4\alpha + 6\beta + 3 z}\big( 3z^2 - 4\alpha^2 -12 \alpha \beta -4
\beta^2 - 4\alpha \gamma - 4 \beta \gamma \big) \non\\
&+e^{4\alpha} m^2 f^2 + 12 e^{4\beta}m^2 h^2 \Big] \, ,
\end{align} 
and the superpotential is given by~\cite{Bena:2010gs}
\begin{equation}
\label{superpotential}
W(\phi)=- 3 e^{2\alpha + 2 \beta}\big( e^{2\alpha} + e^{2\beta} + e^{2
  \gamma} \big) - 6 m^2 e^{-3z} \Big[ h(f-2h) - \frac{1}{54} \Big] \, .
\end{equation}
The background fields satisfy the flow equation\footnote{Note that the equations
  for $a=1,2,3$ are equivalent to the the Ricci--flat K\"ahler
  condition for the metric~\cite{Martelli:2009ga}, while the ones for
  $a=5,6$ give the self--duality of the internal form $F_4$.}
\begin{equation} \label{floweq}
\frac{d\phi^a_0}{d \tau} = \frac12 G^{ab}\frac{\partial W}{\partial \phi^b_0}
\end{equation}
and they are given by the CGLP solution~\cite{Cvetic:2000db}
 \begin{align}
 e^{2\alpha_0}&= \frac13 (2 + \cosh 2r )^{1/4} \cosh r \, ,  \label{CGLPbackground}\\
 e^{2\beta_0}&= \frac13 (2 + \cosh 2r )^{1/4} \sinh r \tanh r \, , \non \\
  e^{2\gamma_0}&=  (2 + \cosh 2r )^{-3/4} \cosh^3 r \, , \non\\
 f_0 &=\frac{1- 3 \cosh^2 r}{3^{3/2}\, \cosh^3 r} \, , \non\\
 h_0&=-\frac{1}{2\,3^{3/2}\, \cosh r} \, , \non\\
z_0 &= \frac13 \log ( m^2 H(r) ) \,  ,\non 
 \end{align}
 where the warp factor $H$ is defined by the following integral:
\begin{equation}\label{WarpFactor}
H(r) = \int^{\infty}_r \frac{3 \,\sech^3u\, \tanh u }{(2 + \cosh
  2u)^{3/4}}\, d u=
\sqrt{2}\,\frac{y\, (7- 5y^4)}{(y^4-1)^{3/2}} + 5 \sqrt{2}\,F \Big(
\arcsin(y^{-1})|-1\Big),
\end{equation}
where
\begin{equation} \label{yvar}
y= (2 + \cosh(2r))^{1/4}
\end{equation}
and $F$ is the incomplete elliptic integral of the first kind
\begin{equation}
F(\phi| q)  = \int_0^{\phi} (1 - q \sin^2 (\theta))^{-1/2} d \theta \,
.
\end{equation}
As shown in~\cite{Bena:2010gs}, it is useful to solve for the
following linear combinations of the fields $\xi_a$ and
$\phi_a$
\begin{align}
\txi_a &= (\xi_1 + \xi_2 + \xi_3, \xi_1-\xi_2 + 3\, \xi_3, \xi_1 + \xi_2
-3 \xi_3, \xi_4, \xi_5, \xi_6) \, ,\\
\tphi_a &= (\phi_1 - \phi_2, \phi_1 + \phi_2 -2\, \phi_3, \phi_3,
\phi_4,\phi_5,\phi_6) \, .
\end{align}
The first--order systems of coupled differential equations for the
fields $\txi_a$ and $\tphi_a$ are:
\begin{align}
 \txi_4'&=\frac19 e^{-3(z_0 + \alpha_0 + \beta_0)} m^2 \big(54 \, h_0(f_0-2h_0)
 -1 \big)\, \txi_4  \, , \label{txi4eq} \\
 \txi_1'&=\frac{2}{9} e^{-3(z_0 + \alpha_0 + \beta_0)} m^2 \big( 54\, h_0(f_0-2h_0)
 - 1 \big) \, \txi_4    \, ,\label{txi1eq} \\
\txi_5'&= \frac12e^{\alpha_0 - \beta_0}\txi_6 -2m^2 h_0 e^{-3(z_0 + \alpha_0
  + \beta_0)} \txi_4   \, ,\label{txi5eq}\\
\txi_6'&= 6 e^{-3\alpha_0 + 3\beta_0}\txi_5 - 2e^{\alpha_0 -
  \beta_0}\txi_6 - 2m^2 (f_0 - 4h_0) e^{-3(z_0 + \alpha_0 +
  \beta_0)}\txi_4 \label{txi6eq}  \, ,\\
\txi_3'&= 4 e^{-\alpha_0 - \beta_0 + 2 \gamma_0} \txi_3 + \frac29 e^{-3(z_0 + \alpha_0 + \beta_0)}  m^2
\big(54(f_0 - 2h_0)h_0-1\big) \txi_4  \, , \label{txi3eq}\\
\txi_2' &= 2 \cosh (\alpha_0 - \beta_0) \txi_2 -\frac32
e^{\alpha_0-\beta_0} \txi_1 + \frac32 e^{-\alpha_0 - \beta_0}
(e^{2\alpha_0}- 2 e^{2\gamma_0})\txi_3 \label{txi2eq}\\
&\quad - 36 h_0 e^{-3\alpha_0 + 3
  \beta_0} \txi_5 + e^{\alpha_0-\beta_0} (f_0 - 4h_0)\txi_6  \, , \non
   \end{align}
and
\begin{align}
 \tphi_1'&= -2\cosh (\alpha_0 - \beta_0) \tphi_1 +
 \frac{1}{12}e^{-3(\alpha_0 + \beta_0)} (-3 \txi_1 + 4\txi_2 + 3
 \txi_3)  \, ,\label{tphi1eq} \\
 \tphi_2'&= -4e^{-\alpha_0 - \beta_0 + 2\gamma_0}\tphi_2 - 6\sinh
 (\alpha_0 - \beta_0) \tphi_1 +
 \frac{1}{12}e^{-3(\alpha_0 + \beta_0)} (-3\txi_1 + 7\txi_3)  \, ,\label{tphi2eq} \\
\tphi_3'&= 3 \sinh (\alpha_0 - \beta_0) \tphi_1 + \frac32 e^{-\alpha_0
  - \beta_0 + 2\gamma_0}\tphi_2 + \frac{1}{12}e^{-3(\alpha_0 +
  \beta_0)}(\txi_1 - 3 \txi_3)
   \, ,\label{tphi3eq}\\
\tphi_5'&= \frac{2}{m^2}e^{-3\alpha_0 + 3\beta_0} (-3 m^2 \tphi_6 + 9
m^2 h_0 \tphi_1 + e^{3 z_0}\txi_5) \, ,
 \label{tphi5eq}\\
\tphi_6'&= \frac{1}{6m^2}e^{\alpha_0 - \beta_0} (-3m^2 \tphi_5 + 12
m^2\tphi_6 - 3m^2(f_0 - 4h_0)\tphi_1 + e^{3z_0}\txi_6)
   \, ,\label{tphi6eq}\\
\tphi_4' &= \frac{1}{9}e^{-3(z_0 + \alpha_0 + \beta_0)} \Big(
2e^{3z_0} \txi_4 + m^2 ((1 + 54h_0(2h_0 - f_0))\tphi_2 + 2 \tphi_3 +
\tphi_4 \label{tphi4eq} \\
&\quad + 18(h_0(-3(f_0 - 2h_0)(2\tphi_3 + \tphi_4) + \tphi_5) + (f_0
- 4h_0)\tphi_6)) \Big)   \, . \non
   \end{align}

%%%%%%%%%%%%%%%%%%%%%%%%%%
\subsection{Solutions for $\txi_a$} \label{subsectxi}
%%%%%%%%%%%%%%%%%%%%%%%%%%

The solution for the modes $\txi_a$ has been derived in
\cite{Bena:2010gs}. We first note that the equation for $\txi_4$ can
be easily integrated by using the flow equations~\eqref{floweq}; for
$a=4$ this reads:
\begin{equation}\label{floweqn4}
z_0'(r) =2 e^{-3(z_0 + \alpha_0 + \beta_0)} m^2 \Big( h_0(f_0-2h_0)
 - \frac{1}{54} \Big).
\end{equation}
This shows that $\txi_4$ is proportional to the warp factor:
\begin{equation}
\txi_4 = m^2 H(r) X_4.
\end{equation}
In terms of the radial variable $r$ the solutions for the remaining
modes are
\begin{align}
 \txi_4&=m^2 X_4 H(r) \, ,\label{txisolution}\\
 \txi_1 &=2\, m^2 X_4 H(r) + X_1 \, ,\non \\
\txi_5&= -\csch^2 r \, \sech r  \,\Big( \frac{3\sqrt{3}}{2}m^2 X_4
\int^r \frac{\csch^3 u}{(\cosh 2u + 2
  )^{3/4}} du+ X_5  \Big) \non\\
&- \cosh r \, \coth^2 r
\, \Big(-\frac{3\sqrt{3}}{2}m^2 X_4 \int^r \frac{\cosh 2u \,\csch^3
  u \,\sech^4 u}{(\cosh 2u + 2)^{3/4}}du+ X_6 \Big) \, ,\non\\
\txi_6&= (3 \cosh 2r + 1)\, \csch^2r \, \sech^3 r   \,\Big(
\frac{3\sqrt{3}}{2}m^2 X_4 \int^r \frac{\csch^3 u}{(\cosh 2u + 2
  )^{3/4}}du + X_5  \Big) \non \\
&+(4
\coth r \,\csch r  - 2\cosh r)
\, \Big(-\frac{3\sqrt{3}}{2}m^2 X_4 \int^r \frac{\cosh 2u \,\csch^3
  u\, \sech^4 u}{(\cosh 2u + 2)^{3/4}}du+ X_6 \Big) \, ,\non\\
\txi_3&=- 6 \sinh^4  r \,(\cosh 2r + 2) \Big( m^2 X_4 \int^r \frac{\csch^3
   u \, \sech^4 u}{(\cosh 2u + 2)^{7/4}} du+ X_3 \Big) \,
,\non\\
\txi_2 &=  \sinh r \cosh r \Big[X_2 + \frac32 X_1 \coth r + 9 X_3\,
\sinh^3 r \,\cosh r  +
\frac{4}{3\sqrt{3}}X_5 \,\csch r\,\sech^5 r\non \\
 &+ \frac{4}{3\sqrt{3}}X_6\,(
\coth r- 3 \tanh r)  +m^2 X_4\, \Big( 3 H(r) \coth  r -
\frac{2 \tanh r\, \sech^3 r}{(\cosh 2r + 2 )^{3/4}} \non\\
&+ 2\, \csch r \,\sech^5 r \int^r \frac{\csch^3 u}{(\cosh 2u + 2
  )^{3/4}}du  + 9 \sinh^3 r \cosh r  \int^r  \frac{\csch^3
    u \, \sech^4 u}{(\cosh 2u + 2)^{7/4}} du \non\\
& + 4 (\cosh 2r -2)\,\csch 2r \int^r \frac{\cosh 2u \,\csch^3
  u \,\sech^4 u}{(\cosh 2u + 2)^{3/4}}du \non  \Big) \Big] \, .\non 
   \end{align}
The zero energy condition~\eqref{ZEC} reads
\begin{equation}\label{ZECX2}
X_2 = 0.
\end{equation}
By using the change of
variables~\eqref{yvar} it is easy to show (see
Appendix~\ref{appelliptic}) that the solution can be expressed
in terms of only two integrals, the warp factor $H(r)$ and the Green's
function~\cite{Bena:2010ze}
\begin{align}
G(r) &= \int^r \frac{3\sqrt{3}\,\csch^3 u}{2\,(\cosh 2u + 2
  )^{3/4}} du = \frac{\sqrt{3}}{2}\Bigg[ \frac{3
 y  (y^4-1)^{1/2} }{\sqrt{2}\,(9-3y^4)}  - 3\,2^{-1/2}\, F
 \Big( \arcsin(y^{-1})|1 \Big) \label{GreenFunction} \non\\
&\qquad - \sqrt{2}\,\Pi \Big(-\sqrt{3}; 
 -\arcsin(y^{-1})|1 \Big) - \sqrt{2}\,\Pi \Big(\sqrt{3};
 -\arcsin(y^{-1})|1 \Big) \Bigg] \, ,
\end{align}
where we use the standard definition for the incomplete elliptic
integral of the third kind
\begin{equation}
\Pi (n;\phi | m) = \int_0^{\phi} (1 - n \sin^2 (\theta))^{-1} (1 - m
\sin^2(\theta))^{-1/2} d\theta\, .
\end{equation}

\subsection{Solutions for $\tphi_a$} \label{subsectphi}

We now present the solution for the $\tphi_a$ modes. Here we show the
result in a compact form in terms of the variable $r$ and we relegate to
Appendix~\ref{appelliptic} the involved analytic expressions which are
obtained by explicitly performing the integrations. The
first--order perturbations to the metric modes and fluxes are
\begin{align}
 \tphi_1&= -\frac{1} {\sinh  2r}  \int^r \frac{9\, \coth u \,\csch u
}{2\, (2 + \cosh 2u)^{3/4}} \,\txi_{123} \,du+ \frac{ Y_1}{\sinh 2r}  \,
,\label{tphisolution}\\
\tphi_2 & =- \frac{9 \,\csch^4 r}{4\,(2 +
  \cosh 2r)} \int^r \frac{\sinh u}{(2 +
  \cosh 2u)^{3/4}}\Big[ (15 + 3\cosh 2u) \, \txi_1 - 12 \,\txi_2 - (23
+ 7\cosh 2u)\,  \txi_3 \Big] du \nn \\
&  \qquad -\frac{3}{2 + \cosh2r} \tphi_1  +  \frac{\csch^4 r}{2 +
  \cosh 2r} \, Y_2  \, , \nn  \\
\tphi_3 &= Y_3 -\frac{9}{32}\int^r \frac{\csch^3 u}{(2 + \cosh 2u)^{3/4}}
\Big[ \txi_1 + 3\cosh 2u \, \txi_{123}  + 3\, \txi_3\Big] du -\frac38 \cosh 2r \,\tphi_1 - \frac38 \,\tphi_2 \, , \nn \\
\tphi_5 &= \sinh^3 r\, \tanh^3 r \, \Lambda_5
+\frac12\cosh r\,(5-\cosh 2r) \, \Lambda_6  \, , \nn \\
\tphi_6 & = -\frac14 (3 + \cosh 2r )\sinh r \,\tanh r \, \Lambda_5  +
\frac12 \cosh^3 r\, \Lambda_6 \,  ,\nn
   \end{align}
where we defined
\begin{align}
  \Lambda_5 &= Y_5 +\frac{1}{24} \int^r \Big[ 12 \sinh u \, \txi_5 - (5 -
  \cosh 2u )\coth^2 u\, \csch u \, \txi_6 \Big] \, H(u) \,du  \\
& \qquad +\frac{\sqrt{3}}{8}\int^r \frac{(2 -\cosh 2u )\csch^3 u }{(2 +
  \cosh 2u )^{3/4}}\, \txi_{123}\,du+ \frac{2
  - \cosh 2r }{6\sqrt{3}}\, \tphi_1  \, , \nn \\
\Lambda_6 & =Y_6+\frac{1}{12} \int^r \sech u  \tanh^3  u \Big[3 \, (3 +
\cosh 2u )\, \txi_5 + \cosh^2 u  \, \txi_6 \Big] \, H(u) \,du  \nn \\
& \qquad -\frac{\sqrt{3}}{8}\int^r (2 + \cosh 2u )^{1/4} \sech^3 u \tanh
 u  \, \txi_{123}\,du - \frac{(2
  + \cosh 2r) \tanh^4  r }{6\sqrt{3}}\, \tphi_1  \, ,\nn 
\end{align}
 and we dubbed $\txi_{123}$ the following combination of $\txi_a$
 \begin{equation}
\txi_{123} = 3 \,\txi_1 - 4 \,\txi_2 - 3\,\txi_3\, .\non
\end{equation}
The last mode we solve for is the perturbation to the warp factor
$\tphi_4$. Its integral expression is
\begin{align}
\tphi_4 &= \frac{1}{m^2 H(r)} \int^r \frac{ 6 m^2 \,\csch^3 u\, H(u) }{(2 +
  \cosh 2u)^{3/4}}\, \txi_4 \,du+  \frac{1}{m^2 H(r)} \int^r \frac{3 m^2\,
  \sech^3 u \tanh u \,(\tphi_2 + 2 \tphi_3)}{(2 + \cosh 2u)^{3/4}}\,du
 \nn \\
& \qquad -  \frac{1}{m^2 H(r)} \int^r \frac{3\sqrt{3}\,m^2\, \csch u \,\sech
  u \,(\csch^2 u \, \tphi_5 + 2\,\sech^2 u\, \tphi_6)}{(2+\cosh 2u)^{3/4}}\,du
+  \frac{Y_4}{m^2 H(r)} \, .\label{tphi4integral}
\end{align}
We now briefly explain the procedure we followed in order to obtain
this solution. We firstly solve the
system~\eqref{tphi1eq}--\eqref{tphi4eq} using the Lagrange method of
variation of parameters. While in principle this produce a solution
with an increasing number of nested integrations, we found that
successive integrations by parts reduce the outcome of this method to
the compact form~\eqref{tphisolution}. We note that since the solution
for the $\txi_a$ modes is analytic in the variable $y$, the
aforementioned solution for the $\tphi_a$ modes contain at most single
integrals of the from
\begin{equation} \label{integralform}
\int^y f(u) \mathbf{L}(u)\,du,
\end{equation}
where $\mathbf{L}(y)$ is a combination of incomplete elliptic integrals and $f(y)$ is
a polynomial function of the variable $y$. In this form the
expressions for the modes $\tphi_a$
can be easily evaluated numerically, and thus provide a full
interpolating solution which connects the IR and the UV region.

The space of solutions we solved for is parametrized by twelve
integration constants $X_a$, $Y_a$, of which only ten are physical
since $X_2$ can be eliminated through the zero energy
condition~\eqref{ZECX2} and $Y_3$ corresponds to a rescaling of the
three--dimensional coordinates.
In Appendix~\ref{appelliptic} we show the full solution obtained
after replacing the analytic expressions for the modes $\txi_a$ and
by recasting some of the integrations in terms of incomplete
elliptic integrals. We were not able to further simplify the resulting
solution, but we stress that the crucial improvement that permits to
easily handle numerical evaluation is the absence of nested integration
(as opposed for example to what happens for the anti--D3 case
\cite{Bena:2011hz}).

%%%%%%%%%%%%%%%%%%%%%%%%%%%%%%%%%%
\section{Asymptotic behavior}\label{secasymtotic}
%%%%%%%%%%%%%%%%%%%%%%%%%%%%%%%%%%

In order to impose the desired boundary conditions we need to
calculate the behavior of the solution presented in the previous
section in the small and large $r$ limits. For that we need the
expansions of the elliptic integrals that enter in the
expressions for the $\tphi_a$ modes. In the IR the first terms of the relevant functions are
\begin{align}
F \Big(\arcsin (y^{-1} ) | -1\Big) & = F_0 - \frac{r^2}{2\sqrt{2}
  \,3^{3/4}} + \frac{r^4}{12\sqrt{2}
  \,3^{3/4}}+ \mathcal{O}(r^6) \, ,\label{Fexpansion} \\
\Pi \Big(-\sqrt{3};-\arcsin (y^{-1}) | -1\Big)& = K_1 + \frac{r^2}{4\sqrt{2}
  \,3^{3/4}}- \frac{r^4}{48\sqrt{2}
  \,3^{3/4}} + \mathcal{O}(r^6) \, , \\
\Pi \Big(\sqrt{3};-\arcsin (y^{-1} )| -1\Big) & =  K_2 +
\frac{3^{1/4}\, \log(r)}{\sqrt{2}} - \frac{r^2}{4\sqrt{2}
  \,3^{3/4}} + \frac{r^4}{40\sqrt{2}
  \,3^{3/4}}+\mathcal{O}(r^6) \, , \label{Pexpansion} 
\end{align} 
where in order to keep notation intelligible, we used the
following abbreviations:
\begin{equation}
F_0 = F
 \Big( \arcsin\Big(\frac{1}{3^{1/4}}\Big)|-1 \Big)  \approx 0.7896\,
 ,\quad K_1 = \Pi \Big(-\sqrt{3};-\arcsin \Big(\frac{1}{3^{1/4}}\Big)
 | -1\Big)\approx -0.6142\non\, 
\end{equation}
and $K_2 \sim -0.9102$. We also encounter the constant
\begin{equation}
K(-1) = 2\,\sqrt{\frac{2}{\pi}}\,\Gamma\left(\frac54\right)^2 \approx
1.3110 \, , \non
\end{equation}
where $K$ is the complete elliptic integral of the first
kind\footnote{Defined as $K(q) = F(\frac{\pi}{2}|q)$.}. Finally, we need the expansion of the warp
factor~\eqref{WarpFactor} which is
\begin{equation}\label{Hexpansion} 
H(r) =  H_0 - \frac12\,3^{1/4}r^2 +
\frac{7\,r^4}{4\,3^{3/4}} +
\mathcal{O}(r^6) \, ,  \qquad H_0 = -4\,3^{1/4} + 5\sqrt{2}F_0\approx
0.3187 \, .
\end{equation}
With these expansions we can easily find the IR behavior of the
solution. In order to
match with the UV behavior we only need to perform a
numerical integration to find the expansions of
the integrals that appear as the coefficients of $X_4$ in the solution shown in Appendix~\ref{appelliptic}.

\subsection{Numerical matching}

\begin{figure}[t]
\begin{center}
\includegraphics[scale=0.7]{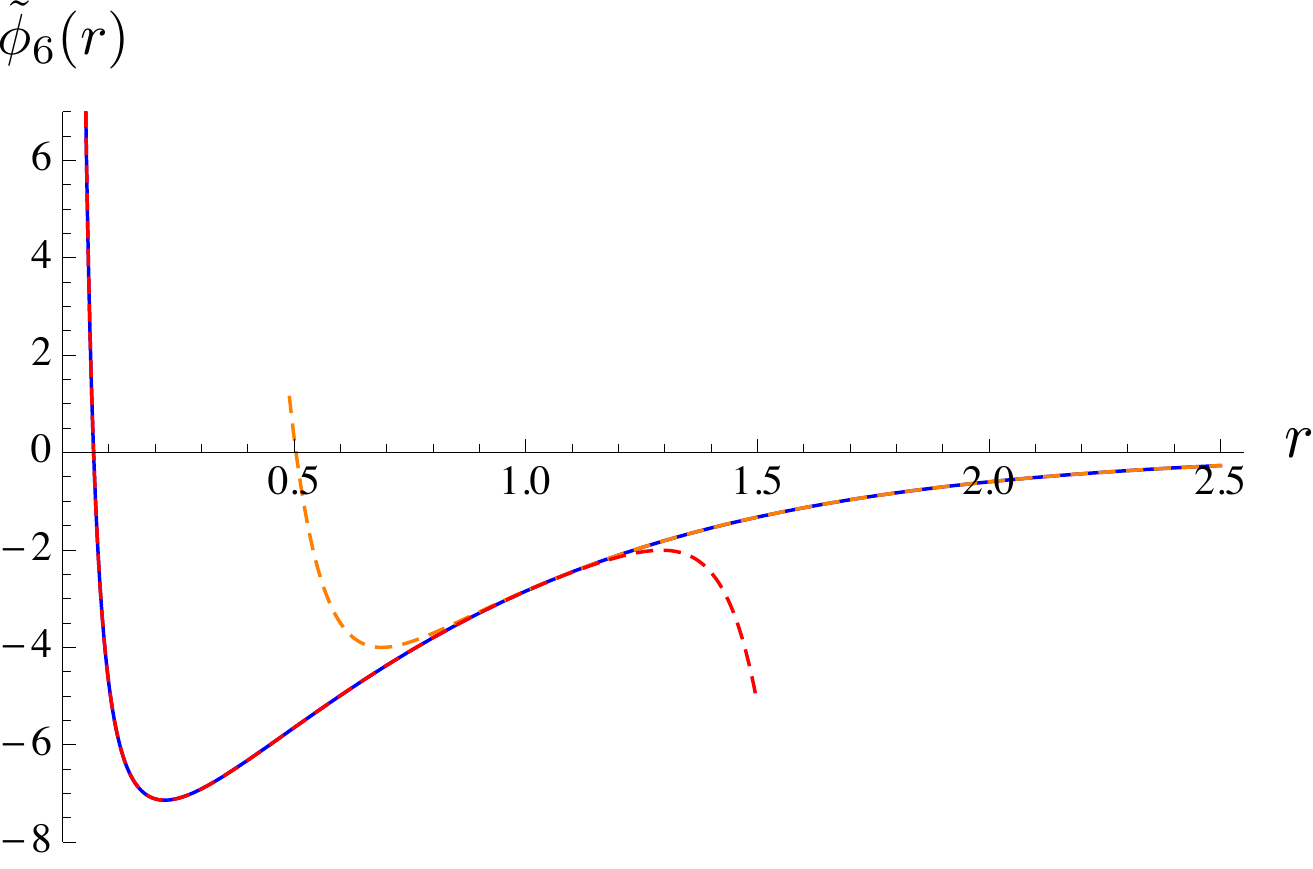} 
\caption{The solution for the mode $\tphi_6$, for $X_2=0$,
  $X_1=X_3=X_5=X_6=1$, $X_4=10$, $Y_a=1$, $m=1$
  (underlying blue solid line). The
  red and orange dashed curves correspond to the IR
  and UV expansions (respectively up to 20 and 15 terms).} \label{plottphi6}
\end{center}
\end{figure}

We now briefly describe the numerical method used to relate the UV and
IR expansions of the integrals that appear in the solution for the $\tphi_a$ modes. They are of the
form~\eqref{integralform}, thus by
using~\eqref{Fexpansion}--\eqref{Pexpansion} we easily get the IR
expansions for the integrands. 
By performing an indefinite integration we therefore get the desired
expansions up to an integration constant which is generically
different in the IR and in the UV. Since these integrals are divergent in the small $r$ limit
but vanish at $r=\infty$, we chose to do the definite integration
in the range $[r, \infty]$; in this way the UV integration constant is
zero and we only need to match in the IR. This can be done up to an
arbitrary precision $p$ by fixing an $r_0$ smaller than the radius of
convergence of the IR series, evaluating the IR expansion
$\mathcal{S}$ of the indefinite
integral at $r_0$ up to the appropriate order $n$ and then fixing a constant
$k$ such that
\begin{equation}
\left| \, \mathcal{S}_n (r_0) + k - \int_{r_0}^{\infty} f(u) \,\mathbf{L}(u)\,du\, \right| <
10^{-p}.
\end{equation}
We kept a precision of $p \approx 10$, which we found enough for our
purposes. As a check, we can verify that the expansions obtained in
the aforementioned way approximates well the numerical solution for
small and large $r$, as shown in Figure~\ref{plottphi6} for one of the
perturbation modes.

\subsection{Infrared expansions}\label{infraredexpansions}

We now show the IR expansions of the modes $\tphi_a$, focusing on the
singular behavior which is needed in order to impose boundary
conditions in Section~\ref{M2bc}. These
  expansions appeared already in~\cite{Bena:2010gs} and apart from
  making sure our results are fine in the present paper we relate
  $Y^{IR}$ to $Y^{UV}$, a crucial step in order to try and write the
  backreacted state at linearized order for all radii.
Here the integration
constants $X_a$ and $Y_a$ are those appearing in the analytic solution shown
in section~\ref{subsectxi} and~\ref{subsectphi} and we defined the $\tilde Y_a$ as
\begin{equation} \label{numcoeff}
\tilde Y_a = Y_a +  \, m^2 \, X_4 \, k_{\tphi_a}  \, ,
\end{equation}
\begin{equation} \nn
k_{\tphi_1} = 7.45479 \, ,\quad k_{\tphi_2} = 0.301287 \, , \quad
k_{\tphi_  3} = 0.112188 \, , \quad k_{\tphi_  5} = 0.576358 \, ,\quad k_{\tphi_  6} = -0.00504419 \, .\end{equation}
The constants $k_{\tphi_a}$ are obtained with the numerical procedure
outlined in the previous subsection. We also impose the zero energy condition~\eqref{ZECX2} and so in what follows we set
$X_2=0$. With these remarks and notations in mind, we now provide the IR
expansions for the first--order perturbation modes
\allowdisplaybreaks{
\begin{align}
\tphi_1 & = \frac{1}{4\, 3^{3/4} r^2} \bigg[ -27 \, X_1 - 
16 \sqrt{3} \, (X_5 + X_6) - 30 \,m^2\,H_0 \,  X_4 \bigg] \\
& \quad + \frac{1}{12 r} \bigg[ 6 \tilde Y_1 + \sqrt{2} \, 3^{1/4} (-45
\sqrt{3}\, X_1 - 162\sqrt{3}\, X_3 + 40\, X_5 + 112\, X_6) K(-1) \bigg] \nn \\
& \quad +  \frac{1}{12\cdot 3^{3/4}}\Big( 189\, X_1 - 80 \sqrt{3} \, ( X_5 +
  X_6) + 6 \, m^2 \, X_4 (83\,H_0-3^{1/4} \,33)\Big) +
  \mathcal{O}(r) \, , \non\\
\tphi_2 & = \frac{1}{r^4}\bigg[\frac{\tilde
  Y_2}{3}+\frac{4}{105}\,H_0\,( 63\, X_1 +
432\,X_3-7\sqrt{3}\,(5\,X_5+11\,X_6)) +\frac{63\,3^{1/4}}{5}\, X_1 +
\frac{2808}{35}\,3^{1/4}\, X_3\non\\
&\quad - \frac{72}{5}\,3^{3/4}\,X_6 \bigg]  + \frac{1}{r^2}\bigg[ -\frac{291}{20}\,3^{1/4}\, X_1  -
\frac{3744}{35}\, 3^{1/4}\,X_3 +
2\,3^{3/4}\,X_5+\frac{106}{5}\,3^{3/4}\,X_6 \\
&\quad +H_0\,\Big(-\frac{16}{5}\, X_1 - \frac{768\, X_3}{35} +
\frac{3\,3^{1/4}\,m^2\,X_4}{2}+
\frac{16\,(5\,X_5+11\,X_6)}{15\sqrt{3}}\Big) -\frac{4\,\tilde Y_2}{9} \bigg] \nn\\
&\quad -\frac{1}{12 r}\bigg[ 6 \, \tilde Y_1-\sqrt{2}\,3^{1/4} (45\sqrt{3}\,
X_1 + 162\sqrt{3}\,X_3 - 40\, X_5 -112\, X_6 ) K(-1) \bigg] \nn \\
&\quad +\frac{137\,3^{1/4}\, X_1}{25} +\frac{12792\,3^{1/4}\,X_3}{175} +
6\sqrt{3}\,m^2\,X_4-\frac{1}{675}(225\,3^{3/4}\,X_5+9081\,3^{3/4}\,X_6-
205\tilde Y_2)\nn\\
&\quad +H_0 \Big(\frac{164}{75}\,X_1 +
\frac{2624}{175}\,X_3-\frac{23\,3^{1/4}\,m^2\,X_4}{2}
-\frac{164\,(5\,X_5+11\,X_6)}{225\sqrt{3}}\Big) +\mathcal{O}(r) \, , \non\\
\tphi_3 & = \frac{1}{r^4}\bigg[ \frac{H_0}{70}\Big(-63\,X_1 - 432\,X_3 + 7\sqrt{3}(5\,X_5 +
11\,X_6)\Big) -\frac{189\,
  3^{1/4}\,X_1}{40}-\frac{1053}{35}\,3^{1/4}\,X_3\\
& \quad + \frac{27\,3^{3/4}\,X_6}{5} -\frac{\tilde Y_2}{8}\bigg]+\frac{1}{r^2}\bigg[ \frac{2H_0}{105}  \Big( 63\,X_1 + 432\, X_3
  -7\sqrt{3}(5\,X_5 + 11\,X_6)\Big)\nn\\
&\quad +\frac{237\,3^{1/4}\,X_1}{40}+
\frac{1404\, 3^{1/4}\, X_3}{35} - \frac{3^{3/4}}{10}(5\,X_5  +77\, X_6) +
  \frac{\tilde Y_2}{6}  \bigg]  \nn\\
&\quad + \frac{61}{160} 3^{1/4} \, X_1-\frac{1167}{100} 3^{1/4} \,
X_3-\frac{67}{16} \sqrt{3} m^2 \, X_4-\frac{671 \, X_5-1033 \, X_6}{120\, 3^{1/4}}
-\frac{41}{360}\,\tilde Y_2+\tilde Y_3\nn \\
&\quad + \frac{F_0}{630\sqrt{2}}\Big( -10836\,X_1 - 69444\, X_3 +
6300\,3^{1/4}\,m^2 \,X_4 + 7\sqrt{3}\,(725\,X_5+1847\,X_6)\Big) \nn \\
&\quad + \frac{K_1+K_2}{4\sqrt{2}}\Big( 21\,X_1 -
4\sqrt{3}(X_5+X_6)\Big) + 30 \sqrt{2}\,3^{1/4}\, m^2\,X_4
\,F_0\log r  \nn\\
&\quad + \frac18\Big( 21\, 3^{1/4}\,X_1- 210\, \sqrt{3}\,m^2\,X_4 -
4\,3^{3/4}\,(X_5+X_6)\Big)\log r + \mathcal{O}(r)\, ,\nn\\
\tphi_5
&=\frac{1}{5}\Big(-12\,3^{3/4}\,X_1-81\,3^{3/4}\,X_3+18\,3^{1/4}\,X_6+10\,\tilde
Y_6\Big) \\
&\quad
-\frac{H_0}{60}\Big(21\sqrt{3}\,X_1+4\,(27\sqrt{3}\,X_3+10\,X_5+4\,X_6)\Big)+
\mathcal{O}(r^3) \nn \, , \\
\tphi_6 & =
\frac{H_0}{12\,r^2}\bigg[\sqrt{3}\,H_0\,m^2\,X_4-2\,(X_5+X_6)\bigg] - H_0\,\Big(\frac{7\sqrt{3}\,X_1}{80}+\frac{9\sqrt{3}\,X_3}{20}+\frac{7\,m^2\,X_4}{24\,3^{1/4}}+\frac{10\,X_5+X_6}{90}\Big)\nn \\
&\quad
-\frac{H_0^2\,m^2\,X_4}{12\sqrt{3}}-\frac{43\,3^{3/4}\,X_1}{80}-\frac{81\,3^{3/4}\,X_3}{20}+\frac{25\,X_5+106\,X_6}{30\,3^{3/4}}+\frac{\tilde
  Y_6}{2} + \mathcal{O}(r) \, , \\
\tphi_4 & =
\frac{1}{r^2}\Big[-\frac34\,\,3^{1/4}\,H_0^2\,m^2\,X_4-\frac{1}{H_0}\Big(\frac{207\sqrt{3}\,X_1}{40}+\frac{2403\sqrt{3}\,X_3}{70}-\frac{54}{5}\,X_6+\frac{\tilde
  Y_2}{8\,3^{3/4}}-3^{3/4}\,\tilde Y_6\Big) \nn\\
&\quad -\frac{33}{40}\,3^{1/4}\,X_4
-\frac{333}{70}\,3^{1/4}\,X_3-\frac{5\,X_5-X_6}{5\,3^{1/4}}\bigg]
+\mathcal{O}(r^0) \, .
\end{align} }

\subsection{Ultraviolet expansions}\label{ultravioletexpansions}

Here we show the leading terms in the UV expansions of the
perturbation modes $\tphi_a$
\begin{align}
\tphi_1 & = -\frac{27\, X_3}{2^{1/4}} e^{-r/2} + 2e^{-2r}\, Y_1- e^{-5r/2}\,2^{7/4}\Big( 27\, X_1  + 81\, X_3 - 16\sqrt{3}\, X_6\Big)  \\
&\quad +\frac{1}{20\,2^{1/4}} e^{-9r/2}\Big(3267\,X_3-1024\sqrt{3}\,X_6\Big)+\mathcal{O}(e^{-6r})\, ,\nn\\
\tphi_2 & = -\frac{63\, X_3}{10\,2^{1/4}}e^{3 r/2} 
-\frac{52 569\, X_3}{280\,2^{1/4}}e^{-5r/2}-12e^{-4 r} \,Y_1
+\mathcal{O}(e^{-9r/2})\, ,\\
\tphi_3 &=  Y_3 -\frac{3 Y_1}{8}+\frac{81 \,X_3}{20\, 2^{1/4}
  }e^{3r/2} -\frac{29079 \,X_3}{560\, 2^{1/4}} e^{-5r/2}+\frac{15}{4} e^{-4 r} \, Y_1 +\mathcal{O}(e^{-9r/2})\, ,\\
\tphi_5 &= \frac{Y_5 - Y_6}{8}e^{3r} - \frac{9\, (Y_5 - Y_6)}{8}e^r  +
  \frac{1}{24} e^{-r} \Big( -8\sqrt{3}\, Y_1 + 117 \, Y_5 + 27\, Y_6
  \Big) \label{phi5UV}\\
&\quad - 38\,2^{3/4} \sqrt{3}e^{-3 r/2}\, X_3
+\frac{1}{72}e^{-3r}\Big(168\sqrt{3}Y_1-9 \,(111\,Y_5+Y_6)\Big)\non \\
& \quad +\frac{2}{195}2^{3/4} e^{-7r/2}(-3348\sqrt{3} X_1 +1323\sqrt{3}X_3+6160\,X_6) +\mathcal{O}(e^{-5r})\, ,\nn\\
\tphi_6 &= -\frac{Y_5 - Y_6}{16}e^{3r} - \frac{3\, (Y_5 - Y_6)}{16}e^r  +
  \frac{1}{144} e^{-r} \Big(- 8\sqrt{3}\, Y_1 + 117 \, Y_5 + 27\,Y_6
  \Big) \label{phi6UV}\\
&\quad - 5\,2^{3/4} \sqrt{3}\,e^{-3 r/2} \, X_3 +\frac{1}{48}
e^{-3r}\Big(8\sqrt{3} Y_1 -51\,Y_5+3\,Y_6)\Big)\non\\
&\quad +\frac{1}{585} 2^{3/4} e^{-7r/2} \Big(-1188\sqrt{3} X_1 + 243 \sqrt{3} X_3 +2320\, X_6)+\mathcal{O}(e^{-5r}) \, ,\nn\\
\tphi_4 & = \frac{3\,Y_4^{UV}}{16\,2^{3/4} m^2}\,e^{9r/2} + \frac{27
  \, Y_4^{UV}}{26\,2^{3/4} \, m^2}\,e^{5 r/2} -
\frac{27\,X_3}{10\,2^{1/4}}\,e^{3r/2} +
\frac{350271\,Y_4^{UV}}{182872\,2^{3/4}\,m^2}\,e^{r/2}  \\
&\quad + \frac{3\,Y_1}{2}-2\,Y_3 -2\sqrt{3}\,Y_5 + 2\sqrt{Y_6} -
\frac{324}{325}\,2^{3/4}\,X_3\,e^{-r/2}+\frac{484605\,Y_4^{UV}}{298792\,2^{3/4}\,m^2}\,e^{-3r/2}
\nn\\
&\quad
-\frac{24}{13}(Y_1-6\sqrt{3}\,Y_6)\,e^{-2r}-\frac{11957859009\,X_3}{28155400\,2^{1/4}}\,e^{-5r/2}
+\frac{7978373883\,Y_4^{UV}}{21130570240\,2^{3/4}\,m^2}\,e^{-7r/2}  +\mathcal{O}(e^{-4r})\, .\non
\end{align} 
From these expansions we can extract the UV behavior of the fields
$\tphi_a$, which is important to understand the holographic
physics. For this purpose we have to relate our radial variable
$r$ to the standard $AdS$ coordinate $\rho_{AdS}$ as
\begin{equation}
\rho_{AdS} \sim e^{3r/2} \, .
\end{equation}
A discussion of the holographic behavior can be found
in~\cite{Bena:2010gs}, where it was shown that the integration
constants $X_a$ and $Y_a$ are paired into normalizable and
non-normalizable mode. In order to be self--contained we tabulate in
Table~\ref{UVmodetable} (which is adapted from~\cite{Bena:2010gs}),
the leading terms coming from each modes. Note that since we obtain
the asymptotic behavior from an analytic solution, we can relate the
integration constants of~\cite{Bena:2010gs} to the IR singular
behavior of the same modes. In particular, one can explicitly check if an IR
regularity condition on one integration constant is compatible with
the absence of the respective
non--normalizable mode in the UV. 
We will come back on this point in the next section. 
In the following table $\Delta$ is the dimension of
the local operator $\mathcal{O}$ holographically associated to the two supergravity
modes whose asymptotic is $\rho_{AdS}^{-\Delta}$ (dual to the vacuum
expectation value of $\mathcal{O}$) and $\rho_{AdS}^{\Delta-3}$ (dual
to a deformation of the action $\delta S=\int
d^{3}x\,\mathcal{O}$). Also, the combination which appears at dimension
$\Delta = 7/3$ is the linear combination of $Y_1$, $Y_5$ and $Y_6$
which appears in the corresponding terms in~\eqref{phi5UV} and
\eqref{phi6UV}. 
\begin{table}[h]
\begin{center}
\begin{tabular}{|c|c|c|c|c|c}\hline
dim $\Delta$ & non-norm/norm & integration constants \\ \hline
6 & $\rho_{AdS}^3/\rho_{AdS}^{-6}$ & $Y_4/X_4$  \\\hline
5 &$\rho_{AdS}^2/\rho_{AdS}^{-5}$ & $Y_5-Y_6/X_5-X_6$ \\\hline
4 & $\rho_{AdS}/\rho_{AdS}^{-4}$  & $X_3/Y_2$ \\\hline
3 & $\rho_{AdS}^0/\rho_{AdS}^{-3}$  & $Y_1+Y_3/X_2$ \\\hline
$\frac73$ & $\rho_{AdS}^{-2/3}/\rho_{AdS}^{-7/3}$  & $Y_5+Y_6+Y_1/X_5+X_6$  \\\hline
$\frac53$ &$\rho_{AdS}^{-4/3}/\rho_{AdS}^{-5/3}$  & $Y_1/X_1$ \\\hline
\end{tabular}
\caption{The UV behavior of all fourteen modes for the
  $SO(5)$-symmetric deformation around the CGLP solution, extracted
  from the asymptotic of our analytic solution.}\label{UVmodetable} 
\end{center}
\end{table}

%%%%%%%%%%%%%%%%%%%%%%%%%%%%%%%%%%%%%%
\section{Charges and M2--branes}\label{secchargeM2brane}
%%%%%%%%%%%%%%%%%%%%%%%%%%%%%%%%%%%%%%

The space of solutions we solved for in the previous sections
should contain the linearized perturbation of the warped Stenzel space
due to the presence of a stack of smeared anti--M2 branes placed at the tip of the geometry. This configuration was studied in the probe approximation
in~\cite{Klebanov:2010qs} and corresponds in the dual gauge theory to
a metastable supersymmetry breaking state. In order to identify the backreacted
solution, we need to impose the correct boundary conditions associated to the presence of the
anti--branes at the tip.
In this section we start by discussing the standard notions of charge in the Stenzel background
 (see for example~\cite{Gukov:1999ya,Aharony:2009fc,Hashimoto:2011aj})
 and as a warmup we identify the BPS perturbation of the CGLP solution
 ascribed to the presence of M2 branes. The anti--M2 brane perturbation will be discussed in the next section.\\

In the Stenzel background we can define a ``running'' M2 charge by integrating $\star_{11} G_4 $ on a 7--dimensional
section $\mathcal{M}_r = V_{5,2}$ of the transverse cone at a fixed
$r$
\begin{equation}\label{M2charge}
Q_{M2}(r) = \frac{1}{(2\pi l_p)^6} \int_{\mathcal{M}_r} \star_{11} G_4  \, ,
\end{equation}
where $l_p$ is the Planck length in eleven dimensions.
We can also integrate $G_4$ over the 4--sphere which has a
finite size at the tip and define the quantity
\begin{equation}
q (r) = \frac{1}{(2\pi l_p)^3}\int_{S_4} G_4 \, .
\end{equation}
For the parametrization~\eqref{metric}-\eqref{fluxes} and for the CGLP
background we find from~\eqref{floweqn4} (see also Appendix~\ref{appforce})
\begin{align}
Q_{M2}^0 (r) &= -\frac{ 6\,2^{11}\,m^2\, \text{Vol}_{V_{5,2}}}{3^4\,(2\pi l_p)^6} \Big( h_0  (f_0-2 h_0)
  -\frac{1}{54}\Big) \, , \label{QM2stenzel}\\
q^0 (r) &=  -\frac{16\pi^2 m}{(2\pi l_p)^3} h_0(r) \, , \label{qM2stenzel}
\end{align}
where $\text{Vol}_{V_{5,2}} =
27\pi^4/128$~\cite{Bergman:2001qi}. Substituting the zeroth--order
solution~\eqref{CGLPbackground} we find
\begin{equation}
Q_{M2}^0 (r)=\frac{ \,m^2\,\tanh^4 r }{108\,\pi^2\, l_p^6}
\, , \quad q^0(r)=\frac{m\,\sech r }{3\sqrt{3}\,\pi\,l_p^{3}} \, ,
\end{equation}
which is the known result for the CGLP solution~\cite{Klebanov:2010qs}. 

\begin{figure}[t]
\begin{center}
\includegraphics[scale=0.7]{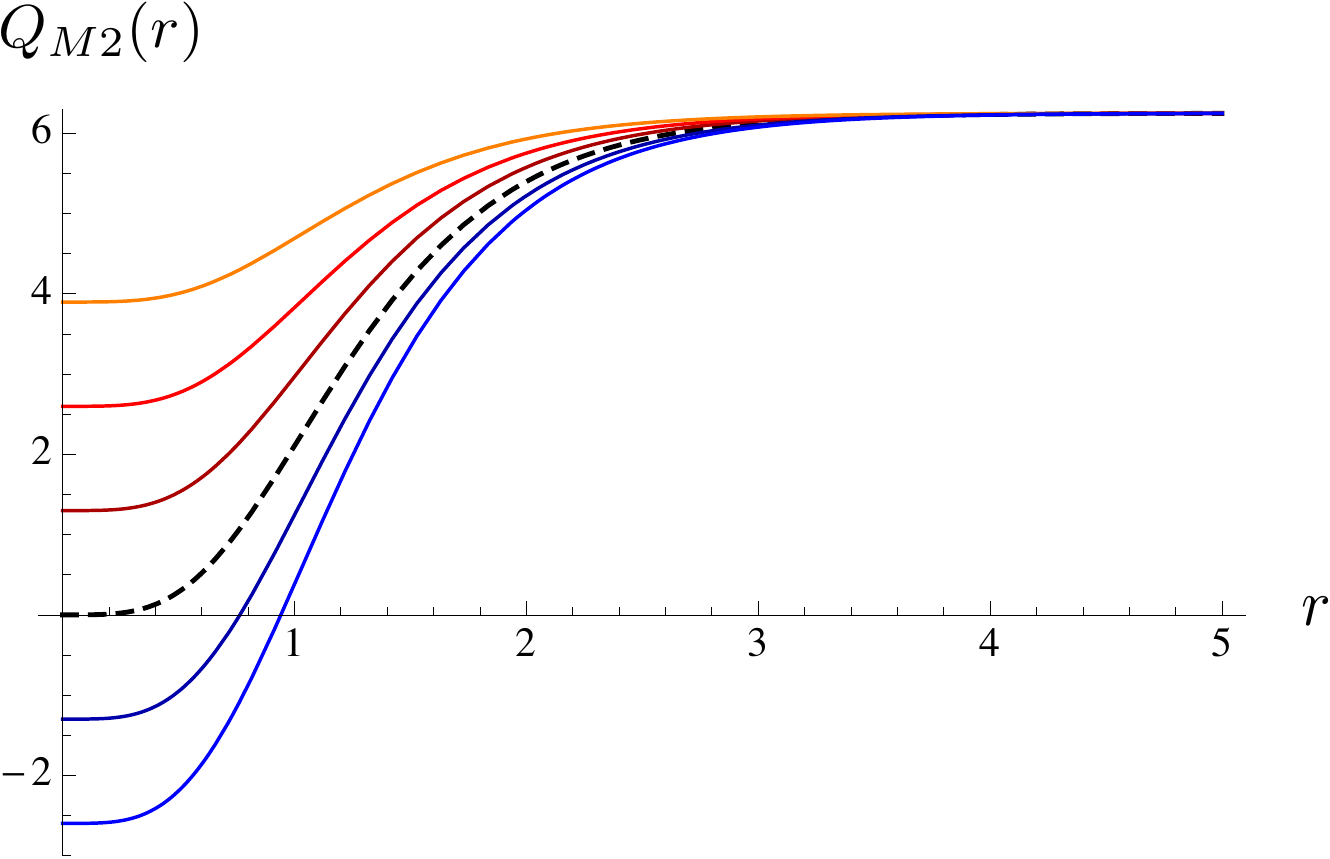}
\caption{The profile of the M2 charge $Q_{M2}$ for the BPS
  perturbation, for
  different values of the parameter $Y_6$. The black dashed line is
  the zeroth--order solution ($Y_6=0$). Note that at the linearized
  level the perturbations vanish in the UV.} \label{plotqbps}
\end{center}
\end{figure}

We now want to calculate the first--order
corrections to these charges from  the first--order perturbation of the Stenzel
geometry. The simpler case is the BPS one, where a
stack of M2--branes smeared over the $S^4$ is placed at the origin $r=0$. The perturbation on
the geometry should still preserve supersymmetry, so we are forced to
set $X_a = 0$, $a =1,\dots, 6$ since the ``conjugate--momenta''
$\txi_a$ are the modes that parametrize the supersymmetry
breaking. Note that in this case the solutions for the modes $\tphi_a$
are given by the homogenous solutions of the coupled system of
ODE's~\eqref{tphi1eq}--\eqref{tphi4eq} and they are easily found by
setting $\txi_a = 0$ in~\eqref{tphisolution}. The perturbation due to
the presence of M2 branes at the tip is found by imposing the following
conditions on the $Y_a$ integration constants: $Y_1 = Y_2 = 0$  to
cancel IR divergencies in $\tphi_1$ and $\tphi_2$, $Y_4=Y_5-Y_6 = 0$ to cancel the divergent terms in the UV
expansions of $\tphi_4$ and $\tphi_5$, and finally $Y_3=0$ to fix the
freedom of rescaling the three--dimensional coordinates. The first--order
perturbation to~\eqref{QM2stenzel} is proportional to $h_0 (\tphi_5 - 4
\tphi_6) + f_0 \tphi_6$
and at the linearized level the running M2 charge is given by
\begin{equation}
Q_{M2} = Q_{M2}^0 + Q_{M2}^1 = \frac{m^2\,l_p^{-6}}{108\,\pi^2}\bigg[\tanh^4 r +6\sqrt{3}\,Y_6\Big(1- \tanh^4 r\Big)
\bigg] \, .
\end{equation}
The profile of the charge is shown in Figure~\ref{plotqbps} for different
values of the constant $Y_6$.
The asymptotic behavior is the following
\begin{equation}\label{chargeM2IRUV}
Q_{M2}^{IR} = \frac{m^2\,l_p^{-6}}{6\sqrt{3}\,\pi^2}\,Y_6+ \mathcal{O}(r^4) \, ,\qquad
Q_{M2}^{UV}  =\frac{m^2\,l_p^{-6}}{108\,\pi^2}+ \mathcal{O}(e^{-2r}) \, .
\end{equation}
The integral of $G_4$ over the four--sphere is given by the behavior
of the mode $h\sim\tphi_6$ and thus
\begin{equation}
q = q^0  + q^1 = \frac{m\,l_p^{-3}}{3\sqrt{3}\pi} \Big(1 - 3\sqrt{3}
\,Y_6\Big)\, \sech r \, .
\end{equation}
We see that $q$ vanishes at infinity while in the IR it approaches a
constant value 
\begin{equation}\label{qIRfirstorder}
q^{IR}= \frac{m\,l_p^{-3}}{3\sqrt{3}\pi}\Big(1- 3\sqrt{3}
\,Y_6\Big) + \mathcal{O}(r^2)\, , \qquad q^{UV} = \mathcal{O}(e^{-r}) \, .
\end{equation}
We will denote $\tilde M = q^0 (0)$ the number of $G_4$ flux units
through the non--vanishing $S^4$ at the tip.
Note that for the zeroth-order solution we have
\begin{equation}\label{qQzerothorder}
\frac{(\tilde M)^2}{4} = Q_{M2}^{UV} \, .
\end{equation}
At first--order, we expect a term related to the explicit brane charge
in the IR; in fact, we easily see from~\eqref{qIRfirstorder} that our
solution satisfies
\begin{equation}\label{qQfirstorder}
\frac{(q^{IR})^2}{4} = 
 Q_{M2}^{UV} -
  Q_{M2}^{IR} \, ,
\end{equation}
which indeed reduces to~\eqref{qQzerothorder} when
$Y_6=0$, which corresponds to having no
regular M2--branes\footnote{Note that equation~\eqref{qQfirstorder} is just the
  standard relation introduced in~\cite{Gukov:1999ya}. The quantities of
equation (2.16) of that reference are $\Phi = Q_{M2}^{UV}$, $N=Q_{M2}^{IR}$ and $\int
G\wedge G = \frac{(q^{IR})^2}{4}$ (see
also~\cite{Hashimoto:2011nn}).}.
Allowing a nonzero $Y_6$ introduces a singularity in the warp factor
$\tphi_4$
\begin{equation}\label{tphi4divIR}
\tphi_4 = \frac{3^{3/4}\,Y_6}{H_0 \,r^2} + \mathcal{O}(r^0) \, ,
\end{equation}
which is the expected divergency due to smeared M2 branes on the $S^4$
at the tip. 

We could have derived these results without
relying on the actual solution for the $\tphi_a$ modes. In fact, the linearized BPS
perturbation can be obtained by simply shifting the fluxes as
follows~\cite{Bena:2010gs}\footnote{By
  shifting the fluxes $f\rightarrow f+2\,c$, $h\rightarrow
  h+\frac{c}{2}$ we can obtain the full nonlinear solution, but this introduce terms proportional to $c^2$ which
  are not seen in our linearized deformation space.}
\begin{align}
\tphi_5 &= 2c \, ,  \\
\tphi_6 &= \frac{c}{2} \, , \non
\end{align}
where $c$ is the number of M2 branes. The M2 charge thus changes in the
following way
\begin{equation}
Q_{M2}^{(1)} =-\frac{m^2\,l_p^{-6}}{2\,\pi^2} \left(h_0 (\tphi_5 - 4
\tphi_6) + f_0 \tphi_6 \right) = 
-\frac{c\,m^2\,l_p^{-6}}{4\,\pi^2} \,f_0 =\frac{c\,m^2\,l_p^{-6}}{6\sqrt{3}\,\pi^2} +\mathcal{O}(r^4)\, ,
\end{equation}
while for the warp factor we have, from~\eqref{tphi4eq}
\begin{equation}
\Delta\tphi_4' = c\, m^2 f_0(r)\, e^{-3(z_0+\alpha_0+\beta_0)} \overset{r
    \rightarrow 0}{\sim}-\frac{2\,3^{3/4}\,c}{H_0\,r^3} \, ,
\end{equation}
from which we get
\begin{equation}
\tphi_4 = \frac{3^{3/4}\,c}{H_0 \,r^2}+\mathcal{O}(r^0) \, ,
\end{equation}
which agrees with~\eqref{tphi4divIR} with the identification $Y_6 =
c$. From this result we can also extract the correct mass/charge normalization
between the warp factor divergency and the charge sourced by the
branes, which will be useful in the next section
\begin{equation}\label{masschargeration}
m^2\,H_0 \,r^2 \, \tphi_4 = 18\cdot3^{1/4} \,\pi^2\,l_p^6\, |Q_{M2}^{IR} |\, .
\end{equation}
 
In the next section, we will turn to the case of interest in which we add a stack of
anti--M2 branes at the tip of the transverse Stenzel space. In this
case the expressions~\eqref{QM2stenzel}, \eqref{qM2stenzel} evaluated
at first--order in perturbation theory will depend on all of the
$X_a, Y_a$ integration constants. However, we have to impose
appropriate regularity conditions for the IR and UV behavior of the
modes $\tphi_a$, and we will see that this fixes all the integration
constants in terms of $X_4$, which is the one responsible for the force
on a probe M2 brane in this background (see section~\ref{subsecforce}), and
$Y_6$. We thus expect the expressions for the charges in the BPS case
to be modified by some pieces proportional to $X_4$. By requiring the
variation in the M2 charge $Q_{M2}^{(1)}$ to be commensurate to the
singularity introduced in the warp factor,
equation~\eqref{masschargeration}, we will derive a
relation which fixes $Y_6$ in terms of $X_4$ and so we will fix all the
integration constants in terms of the number of anti--M2 branes.

%%%%%%%%%%%%%%%%%%%%%%%%%%%%%%%%%%%%%
\section{The anti--M2 brane perturbation}\label{M2bc}
%%%%%%%%%%%%%%%%%%%%%%%%%%%%%%%%%%%%%

In this section we consider the perturbed solution corresponding to a
stack of $\bar N$ anti--M2 branes at the tip of the transverse
geometry.  It was shown in~\cite{Klebanov:2010qs} that in the probe
approximation, for $\bar N/\tilde M \lesssim 0.054$ this configuration is
metastable and will eventually decay into a supersymmetric
configuration in which $\tilde M -1 -\bar N
$ M2 branes are present at the tip \footnote{The units of $G_4$ flux for the susy state are
   then $\tilde M -2$. A way to understand these values is to look
   at~\eqref{qQfirstorder}. We then see that these are the correct
   values so that the charge at infinity is conserved: $Q_{\text{susy}}^{UV} =
   \frac{(\tilde M-2)^2}{4} + \tilde M -1-\bar N - =
   \frac{q^2}{4}-\bar N = Q_{\text{ms}}^{UV}$, where $Q_{\text{susy}}$ and $Q_{\text{ms}}$ are the charges
 for the susy and metastable states.\label{footnoteprobe}}. 
In order to find the supergravity dual to the metastable state, we
will adopt the following strategy. Firstly, we consider the IR
behavior and we allow only for divergencies that are directly sourced by the
anti-branes. Secondly, we demand that the UV non-normalizable
modes described in section~\ref{ultravioletexpansions} are absent, so
that the UV asymptotic is the same as for the original CGLP
background. As we will show, these requirements (together with the
mass/charge normalization discussed in the previous section) provide enough
independent contraints on the deformation space to fix every
integration constants in terms of a single physical quantity, namely
the number of anti--M2 branes present at the tip. We then compute the
relevant charges for the perturbed solution, as well as the explicit
expression for the force felt by probe M2 branes in the backreacted
anti--M2 background.

\subsection{IR and UV boundary conditions}\label{IRandUVbc}

We now proceed to impose regularity conditions on the IR behavior of
the modes $\tphi_a$. We demand that divergencies are zero except for
the singularity in the warp factor $\tphi_4$ which is directly sourced
by the anti--M2 branes.
We first impose the zero energy condition, which amounts to setting
\begin{equation}
X_2 = 0 \, .
\end{equation}
From regularity of $\tphi_1$ we
derive
\begin{align}
X_1 &= -\frac{2}{27} \bigg[ 8\sqrt{3} (X_5 + X_6) +15\,m^2 \,X_4 \,H_0\bigg] \, ,\\
X_5 &= \frac{27\sqrt{3}\,X_3}{20} -
\frac{8\,X_6}{5}-\frac{Y_1}{20\sqrt{2}\,3^{1/4}K(-1)}-\frac{m^2\,X_4\,\big(100\,H_0\,K(-1)+\sqrt{2}\,3^{1/4}k_{\tphi_1}\big)}{80\sqrt{3}\,K(-1)}\, ,\non
\end{align}
while from the singular terms in $\tphi_2$ we derive $X_6$ and $Y_2$ in terms
of $X_3$, $X_4$ and $Y_1$
\begin{align}
X_6 & =\frac{9\sqrt{3}\,X_3}{4} -
\frac{Y_1}{12\sqrt{2}\,3^{1/4}K(-1)}-\frac{m^2\,X_4\,\big(220\,H_0\,K(-1)+\sqrt{2}\,3^{1/4}k_{\tphi_1}\big)}{48\sqrt{3}\,K(-1)}\, ,\\
Y_2 & = \frac{594}{35} \,(3 \,3^{1/4} - H_0)\, X_3-\frac{\sqrt{2} 3^{1/4} (9 \,3^{1/4} + H_0)\,
Y_1}{5\, K(-1)} -\frac{m^2\, X_4}{10\, K(-1)}  \Big[
 \sqrt{2} 3^{1/4} (9\, 3^{1/4} + H_0) k_{\tphi_1}\non\\
&\quad +30 K(-1) k_{\tphi_2} +8 H_0 (153 \,3^{1/4} + 2 H_0) K(-1) \Big]\, .\non
\end{align}
We can check that with these conditions the
other IR divergencies of the modes $\tphi_a$ are automatically
canceled, except for a $\log r$ mode in the IR behavior of the field
$\tphi_3$, which is a perturbation of the metric, which is proportional to $X_4$. It is not clear why one should not be able to kill such
divergent behavior. However, after imposing the previous boundary
conditions, the solution presents an even worse singularity
appearing in the field strength $F_4^2$~\cite{Bena:2010gs}
\begin{equation}
F_4^2 \sim \frac{X_4^2}{r^4} \, ,
\end{equation}
which is quite analogous to the divergence found in the anti--D3
solution~\cite{Bena:2009xk}, with the difference that now the action
is divergent. This behavior is sub--leading with respect to the energy
density associated to the divergency in the warp factor, which is of
order $r^{-6}$.  Note that this is an IR phenomenon insensible to UV
boundary conditions; in fact, the integration constant $X_4$ cannot be
set to zero for the very simple reason that it parametrizes the force
felt by a probe M2 brane~\cite{Bena:2010gs} and thus is indicative of
the presence of anti--M2 branes at the tip. Despite arguments in the
literature, there is not a rigorous proof that shows if this singularity is
acceptable or not\footnote{For the anti--D3 case, it was argued that the
  singularity is an artifact of perturbation theory and will
  disappear in the full backreacted solution
  (see~\cite{Dymarsky:2011pm,McGuirk:2009xx}). However, other works
  pointed out problems in the full solution for antibranes in ISD flux
backgrounds~\cite{Blaback:2010sj,Blaback:2011nz}.}. Given the difficulties in proving this, we will
assume that the singularity is harmless and we will try to see if the
anti--M2 solution develops problems in the UV; if this is not the case,
the solution we find describes the holographic dual of the
conjectured metastable state in the field theory, but clearly a more
detailed study of the IR singularity is required to decide whether this
supergravity solution can be trusted or not.

We now proceed by imposing boundary conditions in the UV, where we
demand that non--normalizable modes in the UV expansions for the modes
$\tphi_a$ are absent. The
first condition is from the $e^{3r/2}$ term in $\tphi_2$, from which
we get
\begin{equation}
X_3 = 0 \, .
\end{equation}
From the divergent term in $\tphi_5$ we get 
\begin{equation}
Y_5 = Y_6 \, ,
\end{equation}
and finally from the term $e^{-2r}$ in $\tphi_1$ we get
\begin{equation}
Y_1= 0 \, . 
\end{equation}
Note that we should allow an $e^{-r}$ term in the fluxes, which is dual to the
dimension $\Delta=7/3$ operator, since it is the charge mode sourced
by the branes.
We thus see that we fixed the ten physically relevant integration
constants in terms of $X_4$ and $Y_6$, which are related respectively
to the force on a probe M2 brane and to the number $\bar N$ of anti--M2
branes placed at the tip~\cite{Bena:2010gs}. 

\subsection{Charges and anti--M2 branes}\label{subschargeantiM2}

In order to relate $X_4$ and $Y_6$ we look at the M2
charge~\eqref{M2charge}. Once all the boundary conditions are imposed,
we get that
\begin{equation}\label{deltacharge}
Q_{M2}^{IR} = \frac{m^2\,\tphi_5(0)}{12\sqrt{3}\,\pi^2\,l_p^6}= \frac{m^2}{6\sqrt{3}\,\pi^2\,l_p^6}\, ( Y_6- \alpha\, m^2\, X_4  ) \, , 
\end{equation}
where the coefficient $\alpha$ is the following combination of the
numerical constants which enters in the expansions for the modes $\tphi_a$
\begin{equation}
\alpha = \frac{H_0 (63\, 3^{1/4} + 22 H_0)}{60\sqrt{3}} - 
 k_{\tphi_6} + \frac{(27 + 
    3^{3/4} \,H_0)\,  k_{\tphi_1}}{
 360 \,\sqrt{2} K(-1)}\sim 0.900178 \, .
\end{equation}
We impose that this variation gives the correct singularity in the
infrared expansion for the warp factor, which is found to be
\begin{equation}\label{divwarpfactor}
\tphi_4 = \frac{3^{3/4}\, (Y_6 - \beta \, m^2\,X_4)}{H_0\,r^2}+\mathcal{O}(r^0)\, ,
\end{equation}
with
\begin{equation}
\beta =\alpha + \frac{H_0^2}{\sqrt{3}}\sim 0.958828 \, .
\end{equation}
 From the mass/charge normalization~\eqref{masschargeration} we thus
 get the following condition
\begin{equation}
 -Y_6+ \alpha \, m^2\, X_4   = Y_6 - \beta \, m^2\,X_4 \, ,
\end{equation}
which results in
\begin{equation}\label{Y6num}
Y_6 = \frac12 (\alpha + \beta) \,m^2\, X_4 = \Big(\alpha +\frac{H_0^2}{2\sqrt{3}}\Big)\,m^2\, X_4 \, .
\end{equation}
If we now plug this relation back into the expression for the
charge~\eqref{deltacharge}, we find the following relation
\begin{equation}\label{deltaQX4}
Q_{M2}^{IR} =-\bar{N} =\frac{H_0^2\,m^4\, X_4}{36\,\pi^2\, l_p^{6}} \, .
\end{equation}
We note that this result does not depend on the UV boundary
conditions. Indeed, although it is not clear from our derivation, it
is easy to show that if we
only impose IR boundary conditions the terms proportional to $X_3, Y_5$ and $Y_1$
that appear in~\eqref{deltacharge} and~\eqref{divwarpfactor} cancel in~\eqref{Y6num}. 

Since $Q_{M2}^{IR}$ is
related to the number $\bar{N}$ of anti--M2 branes placed at the tip,
from~\eqref{deltaQX4} we determine $X_4$ as a function of $\bar N$ and
thus we fix all the integration constants in terms of this
parameter.\\
\begin{figure}[t]
\begin{center}
\includegraphics[scale=0.7]{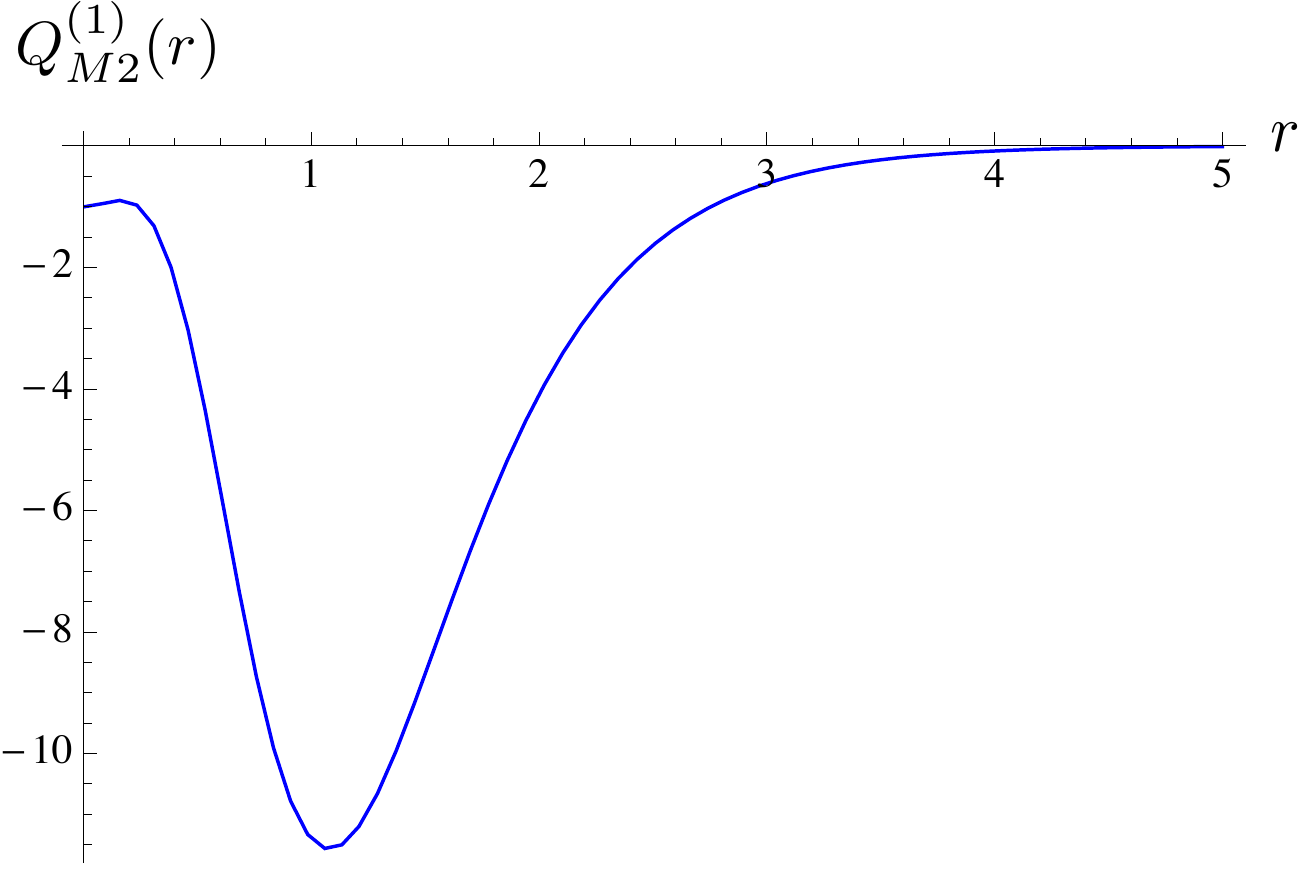} 
\caption{The profile of the first--order M2 charge $Q_{M2}^{(1)}$ for the
  anti--M2 solution, setting $\bar N =1$.} \label{plotantiM2}
\end{center}
\end{figure}

With these results, we can
explicitly compute the charges associated to the anti--M2 brane
perturbation (in Figure~\ref{plotantiM2} we show the profile of the
first--order perturbation to the Maxwell charge $Q_{M2}^{(1)}$). In
particular, the M2 charge $Q_{M2}$ evaluated at a
holographic screen at infinity should be the same for the metastable and the
supersymmetric state. This condition ensures that the metastable state
is a state in the same theory which is dual to the supersymmetric
vacuum.
Unfortunately, we see from~\eqref{chargeM2IRUV} that the perturbation to the M2 charge
vanishes at infinity at the linearized level, and so in our backreacted
solution the value of $Q_{M2}^{UV}$ is fixed. We expect shifts of
this quantity to appear only at second--order in perturbation theory. While
we cannot directly check whether the value of the charge at infinity is
conserved, we can look in the IR and check whether the charges are
perturbed in a consistent way. From the probe computation (see
footnote~\ref{footnoteprobe}), we expect relation~\eqref{qQfirstorder}
to be satisfied. For a supersymmetric domain wall,
this easily follows from the equation of
motion~\eqref{eomG4} and the self--duality of the flux $G_4$, and
indeed we found that the BPS perturbation considered in
section~\ref{secchargeM2brane} is consistent with this constraint at
the linearized level. For the non--supersymmetric case, one should be
more careful. It is useful to write the first--order perturbation to the
Maxwell charge in the IR in the following way
\begin{equation} 
Q_{M2}^{IR} =
\frac{m^2\,l_p^{-6}}{2\pi^2}\bigg[\frac{1}{6\sqrt{3}}\left(\tphi_5(0) -
4\,\tphi_6(0)\right) + \frac{2}{3\sqrt{3}}\tphi_6(0)\bigg]\, ,
\end{equation}
from which we derive, at the linearized level 
\begin{equation}\label{chargesingular}
\frac{(q_{IR})^2}{4} = \frac{m^2 \, l_p^{-6}}{\pi^2} \Big[
-\frac{1}{6\sqrt{3}}+\tphi_6 (0)\Big]^2 = Q_{M2}^{UV} - Q_{M2}^{IR} +
\frac{m^2 \, l_p^{-6}}{12\sqrt{3}\pi^2}\Big[\tphi_5(0) - 4\,
  \tphi_6(0)\Big] +\mathcal{O}(X^2)\, .
\end{equation}
After imposing the anti--M2 IR boundary conditions, we find that the
term in the brakets in
the right hand side of the last equation is not zero
\begin{equation}\label{tphi5tphi6}
\tphi_5 (0) - 4\,\tphi_6(0) = H_0\,3^{-1/4}\,m^2\,X_4 \, .
\end{equation}
Indeed, this is the term which gives rise to the singularity in the
filed strength $F_4^2$ that we discussed in
section~\ref{IRandUVbc}. This result is consistent with the fact that
at the linearized level the self--duality of the four--form flux is
spoiled, and we do not expect relation~\eqref{qQfirstorder} to be
satisfied for the
anti--M2 solution. As we discussed in the previous
subsection, it is possible that this result is an artifact of the
perturbation theory. While we cannot address this issue within
our first--order technology, we believe that further investigation is
needed on this problem.

\subsection{The force on a probe brane}\label{subsecforce}

With the results obtained in the previous subsections, we are able to compute explicitly the
coefficient of the force exerted on a probe M2--brane in the anti--M2 backreacted
background, whose functional form has been derived
in~\cite{Bena:2010gs,Bena:2010ze}:
\begin{equation}\label{forcetext}
F_{M2} = -\frac{18 \, X_4 \, \csch^3 r}{(2 + \cosh 2r)^{3/4}} \, .
\end{equation}
Inserting the expression for $X_4$ that we derive
from~\eqref{deltaQX4} we obtain
\begin{equation}\label{forcefinal}
F_{M2} = \frac{648\,\pi^2\,l_p^6\,\bar{N}\,\csch^3 r 
}{m^4\,H_0^2\, (2 + \cosh 2r)^{3/4}} \, .
\end{equation}
This result has to be compared to the one given by the
probe anti--brane potential \`{a} la KKLMMT \cite{Kachru:2003sx}, which is given
in~\cite{Bena:2010ze} and reviewed in Appendix~\ref{appforce}. The
result of this computation is given in~\eqref{appforceM2}. Once we
substitute $d_2$ we see that the two expressions exactly agree. This
is a nontrivial check that our IR boundary conditions are the correct
ones to describe anti--M2 branes in the Stenzel geometry.

\subsection{Asymptotic of the anti--M2 solution}

We now collect the results we obtained for the twelve $(X_a,Y_a)$
integration constants and which determine the anti--M2 solution in
terms of the constant $X_4$, which is fixed in terms of $\bar{N}$
by~\eqref{deltaQX4}
\begin{equation}
X_4 = -\frac{36\,\pi^2\, l_p^6}{m^4\,H_0^2}\,\bar{N} \, .
\end{equation}
For the $X_a$ integration constants we have
\begin{align}
X_1 &= -2\,H_0\,m^2\,X_4\, , \quad X_2 = 0 \, ,\quad X_3 = 0\, ,\label{bcnumX}\\
 X_5 &= \bigg[ 73\, H_0 +
   2\,3^{1/4}\sqrt{\pi}\,k_{\tphi_1}\,\Gamma\left(\frac14\right)^{-2}\bigg]\,
 \frac{m^2 \,X_4}{12\sqrt{3}} \, ,\non\\
X_6 &=\bigg[-55 \, H_0 -
   2\,3^{1/4}\sqrt{\pi}\,k_{\tphi_1}\,\Gamma\left(\frac14\right)^{-2}\bigg]\,
  \frac{m^2 \,X_4}{12\sqrt{3}} \, ,\non
\end{align}
For the $Y_a$ integration constants we have
\begin{align}
Y_1 &= 0 \, ,\label{bcnumY}\\
Y_2 &= \bigg[-\frac45\,H_0 (153\,3^{1/4} + 2\,H_0)-3\,k_{\tphi_2} -
\frac{4}{5}\,3^{1/4}(9\,3^{1/4}+H_0)\sqrt{\pi}\,\Gamma\left(\frac14\right)^{-2}\bigg]\,
m^2 \, X_4 \, ,\non \\
Y_3 &=0\, , \non\\
Y_5 &=Y_6 =  \bigg[\sqrt{3}\,H_0 (63\,3^{1/4}+52\,H_0) - 180
  \,k_{\tphi_6}+ 2\,(27+3^{3/4}\,H_0)\sqrt{\pi}\,k_{\tphi_1}\,
  \Gamma\left(\frac14\right)^{-2}\bigg]\frac{m^2\,X_4}{180} \, .\non
\end{align}
The IR and UV behavior of the backreacted anti--M2 solution, up to the
desired order, can be
read off from the analytic solution presented in
Appendix~\ref{appelliptic} after imposing the boundary
conditions~\eqref{bcnumX},~\eqref{bcnumY}. For the reader's convenience,
we show here the first few terms of the ultraviolet behavior of the solution.
\allowdisplaybreaks{
\begin{align}
\tphi_1 &=\frac{4\,m^2\,X_4}{3K(-1)}\,\bigg[
29\,2^{3/4}\,H_0\,K(-1)+6^{1/4}\,k_{\tphi_1}\bigg]\,e^{-5r/2}\\
&\quad+\frac{16\,m^2\,X_4}{15K(-1)}\bigg[110\,2^{3/4}\,H_0\,K(-1) +
6^{1/4}\,k_{\tphi_1}\bigg]\,e^{-9r/2} \non\\
&\quad +\frac{7\,m^2\,X_4}{3K(-1)}\,\bigg[
29\,2^{3/4}\,H_0\,K(-1)+6^{1/4}\,k_{\tphi_1}\bigg]\,e^{-13r/2}
+\mathcal{O}(e^{-17r/2}) \, , \non\\
\tphi_2 &=\frac{8\,m^2\,X_4}{3K(-1)}\,\bigg[143\,2^{3/4}\,H_0\,K(-1)+4\,6^{1/4}\,k_{\tphi_1}\bigg]\,e^{-9r/2}\\
&\quad-\frac{16\,m^2\,X_4}{5K(-1)}\bigg[8\,H_0\,(153\,3^{1/4}+2\,H_0)\,K(-1)
+\sqrt{2}\,3^{1/4} (9\,3^{1/4}+H_0)\,k_{\tphi_1}+
30\,K(-1)\,k_{\tphi_2}\bigg]\,e^{-6r} \non\\
&\quad +\,m^2\,X_4\,\bigg[ 2816\,2^{3/4}\,H_0+\frac{128\,6^{1/4}\,k_{\tphi_1}}{5K(-1)}\bigg]\,e^{-13r/2}
+\mathcal{O}(e^{-17r/2}) \, , \non\\
\tphi_3 &=-\frac{4\,m^2\,X_4}{9K(-1)}\,\bigg[295\,2^{3/4}\,H_0\,K(-1)+8\,6^{1/4}\,k_{\tphi_1}\bigg]\,e^{-9r/2}\\
&\quad+\frac{6\,m^2\,X_4}{5K(-1)}\bigg[8\,H_0\,(153\,3^{1/4}+2\,H_0)\,K(-1)
+\sqrt{2}\,3^{1/4} (9\,3^{1/4}+H_0)\,k_{\tphi_1}+
30\,K(-1)\,k_{\tphi_2}\bigg]\,e^{-6r} \non\\
&\quad -\,m^2\,X_4\,\bigg[ \frac{14080}{13}\,2^{3/4}\,H_0+\frac{128\,6^{1/4}\,k_{\tphi_1}}{13K(-1)}\bigg]\,e^{-13r/2}
+\mathcal{O}(e^{-17r/2}) \, , \non\\
\tphi_5 &=\frac{m^2\,X_4}{30} \bigg[\sqrt{3}\,H_0 (63\,3^{1/4}+52\,H_0) - 180
  \,k_{\tphi_6}+ 2\,(27+3^{3/4}\,H_0)\sqrt{\pi}\,k_{\tphi_1}\,
  \Gamma\left(\frac14\right)^{-2}\bigg] \,e^{-r}\\
&\quad-\frac{7\,m^2\,X_4}{90} \bigg[\sqrt{3}\,H_0 (63\,3^{1/4}+52\,H_0) - 180
  \,k_{\tphi_6}+ 2\,(27+3^{3/4}\,H_0)\sqrt{\pi}\,k_{\tphi_1}\,
  \Gamma\left(\frac14\right)^{-2}\bigg]\,e^{-3r} \non\\
&\quad -m^2\,X_4\,\bigg[\frac{48872\,2^{3/4}\,H_0}{585\sqrt{3}}  +
\frac{308\,2^{1/4}\,k_{\tphi_1}}{117\,3^{1/4}\,K(-1)}\bigg]\,e^{-7r/2}\non\\
&\quad+\frac{m^2\,X_4}{6} \bigg[\sqrt{3}\,H_0 (63\,3^{1/4}+52\,H_0) - 180
  \,k_{\tphi_6}+ 2\,(27+3^{3/4}\,H_0)\sqrt{\pi}\,k_{\tphi_1}\,
  \Gamma\left(\frac14\right)^{-2}\bigg] \,e^{-5r}\non \\
&\quad + m^2 X4\bigg[ \frac{301448 \,2^{3/4} \,H_0}{585 \sqrt{3}} +
\frac{131012\,2^{1/4}\,k_{\tphi_1}}{9945\,3^{1/4}\,K(-1)}\bigg]\,e^{-11r/2} 
+\mathcal{O}(e^{-7r}) \, ,\non \\
\tphi_6& =\frac{m^2\,X_4}{180} \bigg[\sqrt{3}\,H_0 (63\,3^{1/4}+52\,H_0) - 180
  \,k_{\tphi_6}+ 2\,(27+3^{3/4}\,H_0)\sqrt{\pi}\,k_{\tphi_1}\,
  \Gamma\left(\frac14\right)^{-2}\bigg] \,e^{-r}\\ 
&\quad-\frac{\,m^2\,X_4}{180} \bigg[\sqrt{3}\,H_0 (63\,3^{1/4}+52\,H_0) - 180
  \,k_{\tphi_6}+ 2\,(27+3^{3/4}\,H_0)\sqrt{\pi}\,k_{\tphi_1}\,
  \Gamma\left(\frac14\right)^{-2}\bigg]\,e^{-3r} \non\\
&\quad -m^2\,X_4\,\bigg[\frac{10516\,2^{3/4}\,H_0}{1755\sqrt{3}}  +
\frac{58\,2^{1/4}\,k_{\tphi_1}}{351\,3^{1/4}\,K(-1)}\bigg]\,e^{-7r/2}
\non \\
&\quad +\frac{m^2\,X_4}{180} \bigg[\sqrt{3}\,H_0 (63\,3^{1/4}+52\,H_0) - 180
  \,k_{\tphi_6}+ 2\,(27+3^{3/4}\,H_0)\sqrt{\pi}\,k_{\tphi_1}\,
  \Gamma\left(\frac14\right)^{-2}\bigg] \,e^{-5r}\non\\
&\quad + m^2 X4\bigg[ \frac{4244 \,2^{3/4} \,H_0}{1755 \sqrt{3}} +
\frac{1466\,2^{1/4}\,k_{\tphi_1}}{3315\,3^{1/4}\,K(-1)}\bigg]\,e^{-11r/2} 
+\mathcal{O}(e^{-7r}) \, .\non \\
\tphi_4& =\frac{4\sqrt{3}\,m^2\,X_4}{65\,} \bigg[\sqrt{3}\,H_0 (63\,3^{1/4}+52\,H_0) - 180
  \,k_{\tphi_6}+ 2\,(27+3^{3/4}\,H_0)\sqrt{\pi}\,k_{\tphi_1}\,
  \Gamma\left(\frac14\right)^{-2}\bigg] \,e^{-2r}\non\\ 
&\quad+\frac{2192\,\sqrt{3}m^2\,X_4}{14365} \bigg[\sqrt{3}\,H_0 (63\,3^{1/4}+52\,H_0) - 180
  \,k_{\tphi_6}+ 2\,(27+3^{3/4}\,H_0)\sqrt{\pi}\,k_{\tphi_1}\,
  \Gamma\left(\frac14\right)^{-2}\bigg] \,e^{-4r}\non\\
&\quad
-\frac{8\,m^2\,X_4}{135K(-1)}\,\bigg[2617\,2^{3/4}\,H_0\,K(-1)+80\,6^{1/4}\,k_{\tphi_1}\bigg]\,e^{-9r/2}
+\mathcal{O}(e^{-6r}) \, .
\end{align}}

\section{Conclusions}

In this paper we constructed the analytic solution for the
twelve--dimensional space of linearized non-supersymmetric
deformations of the warped Stenzel space, consistent with the $SO(5)$
symmetries of the supersymmetric background. Our solution provides an
interpolation between the IR and UV behaviors previously constructed
in~\cite{Bena:2010gs} and it should contain interesting informations
about the dual (2+1)-dimensional gauge theory. In particular, we were
interested in finding the supergravity solution dual to metastable
states, which were conjectured in~\cite{Klebanov:2010qs} to be
described by a stack of anti--M2 branes placed at the tip of the
transverse geometry. We were able to identify this solution by
imposing suitable boundary conditions on the set of twelve integration
constants $(X_a, Y_a)$ that parametrize the full deformation space,
and indeed we showed that this solution is unique and it depends only
on the number $\bar{N}$ of anti--M2 branes placed at the tip. We then
used this solution to compute the force exerted on a probe M2 brane
placed in the anti--M2 backreacted supergravity background and
we showed that it exactly agrees with the calculation \`{a} la KKLMMT~\cite{Kachru:2003sx}
in which one considers the anti--M2 brane as probing the backreacted
geometry of M2 branes on the Stenzel background\footnote{
The agreement between the functional form of these two results has been derived
in~\cite{Bena:2010gs,Bena:2010ze}. Here we are
  computing the full result, including the normalization constant
  and its dependence on $\alpha'$ and $m^2$.}. 

The linearized supergravity solution displays however an IR singularity in
the four-form flux, which leads to a divergent action,
whose nature is still poorly understood. Our analysis shows that this is the only drawback of the
supergravity solution, which otherwise has the desired features to
describe the metastable state in the dual gauge theory. It is thus of
great importance to establish the nature of this singularity. However,
proving if these singularities are acceptable or not is unfortunately beyond the reach of our
first--order technology and clearly more work is needed to address
this problem.

%%%%%%%%%%%%%%%%%%%%%%%%%%%%%%%%%%%%%%%%%%%%%
\vspace{0.5cm}
\noindent {\bf Acknowledgements}:
 \noindent 
I am grateful to Iosif Bena, Gregory Giecold, Enrico Goi, Nick
Halmagyi, Francesco Orsi and especially Mariana
Gra\~na for useful discussions and comments. I am
thankful to Gregory Giecold and  Mariana Gra\~na for helpful observations on  the manuscript.
This work is supported by a Contrat de Formation par la Recherche of
CEA/Saclay and by the ERC Starting Independent Researcher Grant 259133 -- ObservableString. 
%%%%%%%%%%%%%%%%%%%%%%%%%%%%%%%%%%%%%%%%%%%%%

\appendix

%%%%%%%%%%%%%%%%%%%%%%%%%%%%%%%%%%
\section{Analytic solutions}\label{appelliptic}
%%%%%%%%%%%%%%%%%%%%%%%%%%%%%%%%%%

Here we show the analytic solutions for the modes $\txi_a$ which can be
obtained by explicitly performing the integrations that appear in~\eqref{txisolution}
\begin{align}
 \txi_4&=m^2 X_4 H(y) \, ,  \label{txisolution2} \\
\txi_1 &=2 m^2 X_4 H(y) + X_1 \, ,\non \\
\txi_5&= -2\sqrt{2}(y^4-3)^{-1}(y^4-1)^{-1/2} L_5(y)-2^{-1/2}
(y^4-3)^{-1}(y^4-1)^{3/2}L_6(y) \, , \non \\
\txi_6&= 4\sqrt{2}(y^4-3)^{-1}(y^4 -1)^{-3/2}(3 y^4 -5)L_5(y) -
\sqrt{2}(y^4-3)^{-1}(y^4-7)(y^4-1)^{1/2}L_6(y) \, ,\nn \\
\txi_3&= -\frac32 y^4 (y^4 - 3)^2 L_3 (y) \,
, \non \\
\txi_2 &= \frac34 X_1 (y^4 -1) + \frac12 X_2 (y^8 -4y^4 + 3)^{1/2} +
\frac98(y^4-3)^2(y^4-1)L_3(y) + \frac{16}{3\sqrt{3}}(y^4-1)^{-2} L_5(y) \non \\
 &\qquad  -\frac{4}{3\sqrt{3}}(y^4-4)L_6(y) + m^2 X_4 \Big( \frac32 (y^4 - 1)H(y) - 2
\sqrt{2}y^{-3}(y^4-1)^{-3/2}(y^4-3)   \Big) \, , \non 
   \end{align}
where
\begin{align}
 L_5(y) &=X_5 +m^2 X_4 G(y)\, ,\label{Lintegrals}\\
L_6 (y) &=X_6  -m^2X_4 \Big(G(y) + \frac{\sqrt{3}}{2}H(y)\Big)\, \non \\
 L_3(y)&=X_3 +m^2 X_4 \Big(\frac{16\sqrt{2}
  (2y^4-3)}{27y^3(y^4-3)(y^4-1)^{3/2}}+\frac{22
  G(y)}{27\sqrt{3}}-\frac{13H(y)}{27} \Big)\, . 
   \end{align}
We recall that the variable $y$ is defined as
\be
y = (2 + \cosh(2r))^{1/4} \, 
\ee
and the expression for the warp factor $H(y)$ and the Green's function
$G(y)$ are given in~\eqref{WarpFactor} and~\eqref{GreenFunction}.

We now show the expanded form of the solution for the modes $\tphi_a$
in terms of the variable $y$, obtained by replacing the analytic
solutions for the modes $\txi_a$
in~\eqref{tphisolution}. We impose the zero energy condition, so we
put $X_2 = 0$.
{\allowdisplaybreaks  
\begin{align}
\tphi_1 & = \frac{1}{(3- 4 y^4 + y^8)^{1/2}} \Bigg[ Y_1 - \frac{9 y
  X_1}{ \sqrt{2}(y^4 - 3)^{1/2}} 
- \frac{27}{\sqrt{2}} X_3 y
 \sqrt{y^4-3} + \frac{  4\sqrt{2}X_5 y(11-5y^4)}{ (y^4-1)\sqrt{3}(y^4-3)^{1/2}}\non\\[5pt]
&\quad -\frac{8\sqrt{2}X_6 y }{\sqrt{3}(y^4-3)^{1/2}} -
\frac{1}{\sqrt{2}\,3^{3/4}} F\Big( \arcsin(3^{1/4}y^{-1})|-1
  \Big)\Big(45\sqrt{3}X_1 + 162 \sqrt{3}X_3 - 40 X_5 -112 X_6\Big)
  \non \\[5pt]
&\quad + \frac12 m^2 X_4 \bigg(\int^y \frac{3\sqrt{2}(-19-26u^4+13
  u^8) H(u) }{(u^4-3)^{3/2}}du
-\int^y \frac{2\sqrt{6}\sqrt{u^4-3}(11-38u^4+11u^8) G(u)}{(u^4-1)^2}du
\non\\[5pt]
&\quad -\frac{48 y^2 (y^4-3)}{(y^8-4y^4+3)^{1/2}}+ 48\sqrt{3}
E\Big(\arcsin (y^2)|\frac13\Big) -  32\sqrt{3} F\Big(\arcsin(y^2) |\frac13\Big)\bigg)
\Bigg] \, , \\[5pt]
\tphi_2 &=   -\frac{3\tphi_1}{y^4} + \frac{4}{y^4 (y^4-3)^2}\Bigg[
Y_2+ \frac{9 X_1 y \sqrt{y^4-1}}{\sqrt{2}}+\frac{9 X_3   y \sqrt{y^4-1}}{140 \sqrt{2}} (825 - 639 y^4 + 343 y^8 - 49 y^{12})\non\\[5pt]
&\quad  + \frac{4\sqrt{6} X_5 y (5y^4-7)}{(y^4-1)^{3/2}} +\frac47 \sqrt{2} \Big(63 X_1 + 432 X_3 - 7 \sqrt{3} (5 X_5 + 11 X_6)\Big) F\Big(\arcsin(y^{-1}) |-1\Big)\nn \\[5pt]
&\quad - 4
\sqrt{6}X_6 y \sqrt{y^4-1} + 3m^2 X_4\bigg( \int^y \frac{(-171 - 342 u ^4 + 936 u^8 - 546 u^{12}
  + 91 u^{16})H(u) }{12\sqrt{2}(u^4-1)^{1/2}}du\non \\[5pt]
& \quad  -\int^y \frac{(u^4-3)^2 (99-342 u^4+176
  u^8 - 154u^{12} + 77u^{16})G(u)}{6\sqrt{6}(u^4-1)^{5/2}} d u \nn\\[5pt]
&\quad + \frac{2y^2(138-119y^4+14y^8)}{9(1-y^4)}+\frac{16}{3}\text{arccoth}\,y^2 \bigg)\Bigg]\, , \\[5pt]
\tphi_3 & = Y_3 - \frac38( y^4-2)\, \tphi_1 - \frac38\, \tphi_2
+\frac{27 X_1 y \sqrt{y^4-1}}{8\sqrt{2}(y^4-3)}
+\frac{27 X_3 y \sqrt{y^4-1}}{4\sqrt{2}}  +\frac{\sqrt{3} X_5 y (5y^4-7)}{(y^4-1)^{3/2}}  \nn \\[5pt]
& \quad +\frac{1}{8\sqrt{2}}\Big(135 X_1 + 432 X_3 - 8 \sqrt{3} (5 X_5 + 12 X_6)\Big) F\Big(\arcsin(y^{-1}) |-1\Big)\nn \\[5pt]
&\quad +\frac14 \Big(7 \sqrt{3} X_1 - 4 (X_5 + X_6)\Big)G(y)  + 3 m^2 X_4 \bigg[  \int^y
\frac{(-51+9u^4+39u^8-13u^{12})H(u)}{4\sqrt{2}(u^4-3)^{3/2}(3-4u^4+u^8)^{1/2}}du
\non
\\[5pt]
&\hspace{-0.5cm} + \int^y\frac{(-33+125 u^4-79u^8+11u^{12})G(u)}{2\sqrt{6}(u^4-1)^{5/2}}du +\frac{5 y^2}{4-4y^4}-\frac{\sqrt{3}}{4}\text{arctanh}
\left(\frac{y^2}{\sqrt{3}}\right) +
\log\left(\frac{1+y^2}{1-y^2}\right) \bigg] \, .
\end{align}
For the flux perturbation we have
\begin{equation}\renewcommand\arraystretch{2}\begin{pmatrix}
\tphi_5 \\\tphi_6\end{pmatrix} = \begin{pmatrix}
\frac{1}{2\sqrt{2}}(y^4-1)^{-3/2}( y^4-3)^3 & -
\frac{1}{2\sqrt{2}}(y^4-7)(y^4-1)^{1/2}\\
 -\frac{1}{4\sqrt{2}}(y^4-1)^{-1/2}( y^4-3)(y^4+1) &
\frac{1}{4\sqrt{2}}(y^4-1)^{3/2}
\end{pmatrix} 
\begin{pmatrix} \Lambda_5 \\ \Lambda_6\end{pmatrix}
\end{equation}
where
\begin{align}
\Lambda_5 & = Y_5 + \frac{4-y^4}{6\sqrt{3}}\, \tphi_1 +
\frac{3\sqrt{3}X_1 y (7-5y^4)}{5\sqrt{2}(y^4-1)^{3/2}}
-\frac{3\sqrt{3}X_3 y (-72 + 45y^4+5y^8)}{10\sqrt{2}(y^4-1)^{3/2}} \\[5pt]
& \quad   -\frac{10\sqrt{2}X_5 y
}{9(y^4-3)(y^4-1)^{3/2}}+\frac{2\sqrt{2} X_6 y (272 - 279y^4 +
  60y^8)}{45 (y^4-3)(y^4-1)^{3/2}}-\frac{1}{108}\Big(27 X_1 + 44\sqrt{3} (X_5+ X_6)\Big) G(y) \nn \\[5pt]
&\quad +\frac{1}{180}\Big(-54\sqrt{3}X_1 - 297\sqrt{3}X_3 + 40 X_5 +
106 X_6 - 30 X_6 y^4 + \frac{120(X_5 + X_6)}{(y^4-3)^2}\Big)H(y)
\non \\[5pt]
& \quad  +m^2 X_4 \bigg[ \int^y H(u)\left( \frac{76 +
  85u^4-78u^8+13u^{12}}{2\sqrt{6}(u^4-3)^{3/2}(u^8-4u^4+3)^{1/2}} +\frac23
y^3 G(y) + \frac{y^3}{\sqrt{3}}H(y) +\frac{8 y^3 H(y)}{\sqrt{3}(y^4-3)^3}\right)du  \non
\\[5pt]
& \quad +\int^y \frac{(44-163 u^4+82 u^8 - 11 u^{12})G(u)}{3\sqrt{2}(u^4-1)^{5/2}}du+\frac{2\sqrt{3} y^2}{y^4-1} -\frac23
\text{arctanh}\left(\frac{y^2}{\sqrt{2}}\right) - \frac{2}{\sqrt{3}}\log\left(\frac{1+y^2}{1-y^2}\right) \bigg]   \, ,\non\\[5pt]
\Lambda_6 &= Y_6 - \frac{y^4 (-3+y^4)^2}{6\sqrt{3}(-1+y^4)^2}\,
\tphi_1  +\frac{3\sqrt{3} X_1
  y(7-5y^4)}{5\sqrt{2}(y^4-1)^{3/2}}+\frac{3\sqrt{3}X_3y(72-45y^4-5y^8)}{10\sqrt{2}(y^4-1)^{3/2}}
+ \frac{2\sqrt{2}X_5 y (y^4-3)}{3(y^4-1)^{7/2}} \nn \\[5pt]
& \quad +\frac{2\sqrt{2}X_6 y ( 20y^4-33)}{15(y^4-1)^{3/2}} +
H(y)\Big(-\frac{7 \sqrt{3} X_1}{40} - \frac{9 \sqrt{3} X_3}{10} +\frac{X_6 (11 - 5 y^4)}{30} + \frac{X_5 (1 + 6 y^4 - 3 y^8)}{6 ( y^4-1)^2}\Big) \non \\[5pt]
&\quad + m^2 X_4 \bigg[ \int^y
H(u)\left(\frac{y^4(-19-26y^4+13y^8)}{2\sqrt{6}(y^4-1)^{5/2}}+\frac{2y^3(y^4-3)(y^8+3)G(y)}{3(y^4-1)^3}+\frac{
y^3H(y)}{\sqrt{3}}\right)\non\\[5pt]
&\quad  -\int^y \frac{u^4 (u^4-3)^2 (11-38u^4+11u^8)
  G(u)}{3\sqrt{2}(u^4-1)^{9/2}}
+\frac{2y^3(-3-4y^4+3y^8)}{3\sqrt{3}(y^4-1)^3} +
\frac{1}{\sqrt{3}}\log \left(\frac{y^2-1}{-1-y^2}\right) \bigg] \, . \non
\end{align} }

\section{Brane/antibrane potential}\label{appforce}

In this Appendix we review the calculation of the force on a probe
antibrane in the Stenzel background which has been performed
in~\cite{Bena:2010ze}. 
We consider a stack of M2 branes at a position $r=r_0$ in the
transverse geometry and we want to compute the force exerted on a
probe anti--M2 brane placed at the tip $r=0$, due to the backreaction
of the M2 branes. To compute the full backreacted geometry we only
need to add a harmonic function $\delta H(r)$ to the background warp factor
$H_0(r)$~\cite{Grana:2000jj}. This function is given by the Green's function on the warped
Stenzel space~\cite{Pufu:2010ie} and since we are considering smeared branes, we only need
to solve the radially symmetric Laplace equation. The Laplacian is
given by
\begin{equation}
\Delta \delta H  = \frac{1}{\sqrt{G}}\frac{\partial}{\partial x^l}\left(
  \sqrt{G} g^{lm} \frac{\partial \delta H}{\partial x^m}\right) \, ,
\end{equation}
where we are labeling the eleven dimensional coordinates by $x_l$
($l,m=0,\dots,10$) and $G= \det g_{lm}$. From~\eqref{metric}  we easily get
\begin{equation}
\sqrt{G} = -e^{z+3\alpha+3\beta+2\gamma} \, , \qquad g^{rr} =
e^{-z-2\gamma}\, ,
\end{equation}
and so imposing $\Delta \delta H = 0$ we get the following equation
for $\delta H' (r)$
\begin{equation}
e^{3 (\alpha_0 + \beta_0)} \delta H' (r) = \text{const} \, .
\end{equation}
So the two solutions of the Laplace equation are, using~\eqref{CGLPbackground}
\begin{align}
H_{1} &= d_1 \, \\
H_{2} &= d_2\int^r \frac{ \csch^3 u}{(2+\cosh 2u)^{3/4}} dr\,  , \label{appH2}
\end{align}
and we should set $\delta H = H_1$ for $r< r_0$ and $\delta H = H_2 $
for $r >r_0$. The constant $d_1$ is then fixed from the matching
condition $d_1 = H_2(r_0)$ and the constant $d_2$ is related to the
number of M2 branes from~\eqref{M2charge}. In fact, we have
\begin{equation}
N=\frac{1}{(2\pi l_p)^6} \int_{\delta \mathcal{M}}\star_{11} G_4 =
\frac{2^{11}\,m^2\,\text{Vol}_{V_{5,2}}}{3^4\,(2\pi l_p)^6} \, H_2' \, e^{3 (\alpha_0 + \beta_0)}\, ,
\end{equation}
where $\delta \mathcal{M}$ is a small shell around $r_0$. If we use
$H_2$ given by~\eqref{appH2}, the above
equation thus fixes $d_2$ in terms of $N$
\begin{equation}
d_2 = \frac{324\,\pi^2\, l_p^{-6} \,N}{m^2} \, . 
\end{equation}

We now compute the force exerted on the probe anti--M2 brane by
looking at the variation in the potential when we move the stack of M2
branes away from $r=r_0$. For anti--M2 branes, $V_{DBI}=V_{WZ}$ and
since we have
\begin{equation}
V_{DBI} \sim (g_{00}g_{11}g_{22})^{1/2} = e^{-3z} = \frac{1}{m^2\,H} \, ,
\end{equation}
the potential is just proportional to $2 H^{-1}$. Expanding this we
get
\begin{equation}
V =  \frac{2}{m^2\,H} \approx \frac{2}{m^2 H_0}\left(1- \frac{\delta
    H}{H_0}\right) \, ,
\end{equation}
and so we easily get the force:
\begin{align}\label{appforceM2}
F_{M2} &= -\frac{2}{m^2}\frac{\partial V}{\partial r_0} 
\Big\vert_{r=0}  \non\\
&= \frac{1}{m^2\,H_0^2}\frac{2\,d_2 \,\csch^3 r}{(2 + \cosh 2r)^{3/4}} \, .
\end{align}
This result agree with the computation of the force exerted on a probe
M2 brane due to the backreaction of a stack of anti--M2
branes~\eqref{forcefinal}.

%%%%%%%%%%%%%%%%%%%%%%%%%%%%%%%%%%%%%%%%%%%%%%%%%%%%%%%%%%%%
\providecommand{\href}[2]{#2}\begingroup\raggedright\endgroup


\begin{thebibliography}{10}


\bibitem{Intriligator:2006dd}
K.~Intriligator, N.~Seiberg and D.~Shih, ``Dynamical susy breaking in
  meta-stable vacua,'' {\em JHEP} {\bf 04} (2006) 021,
\href{http://arXiv.org/abs/hep-th/0602239}{{\tt hep-th/0602239}}.
%%CITATION = HEP-TH 0602239;%%.



\bibitem{Bena:2006rg}
I.~Bena, E.~Gorbatov, S.~Hellerman, N.~Seiberg, and D.~Shih, ``A note on
  (meta)stable brane configurations in mqcd,''
\href{http://arXiv.org/abs/hep-th/0608157}{{\tt hep-th/0608157}}.
%%CITATION = HEP-TH 0608157;%%.


\bibitem{Klebanov:2000hb}
I.~R. Klebanov and M.~J. Strassler, ``Supergravity and a confining gauge
  theory: Duality cascades and chisb-resolution of naked singularities,'' {\em
  JHEP} {\bf 08} (2000) 052,
\href{http://arXiv.org/abs/hep-th/0007191}{{\tt hep-th/0007191}}.
%%CITATION = HEP-TH 0007191;%%.

\bibitem{Kachru:2002gs}
S.~Kachru, J.~Pearson and H.~L. Verlinde, ``{Brane/flux Annihilation and the
  String Dual of a Non-Supersymmetric Field Theory},'' {\em JHEP} {\bf 06}
  (2002) 021,
\href{http://arXiv.org/abs/hep-th/0112197}{{\tt hep-th/0112197}}.
%%CITATION = HEP-TH 0112197;%%.

\bibitem{DeWolfe:2004qx}
O.~DeWolfe, S.~Kachru and H.~L. Verlinde, ``{The giant inflaton},'' {\em JHEP}
  {\bf 05} (2004) 017,
\href{http://arXiv.org/abs/hep-th/0403123}{{\tt hep-th/0403123}}.
%%CITATION = HEP-TH/0403123;%%.


%\cite{Myers:1999ps}
\bibitem{Myers:1999ps}
  R.~C.~Myers,
  ``Dielectric branes,''
  JHEP {\bf 9912} (1999) 022
\href{http://arXiv.org/abs/hep-th/9910053}{{\tt hep-th/9910053}}.


\bibitem{Bena:2009xk}
I.~Bena, M.~Grana, and N.~Halmagyi, ``{On the Existence of Meta-stable Vacua in
  Klebanov-Strassler},'' {\em JHEP} {\bf 1009} (2010) 087,
  \href{http://arXiv.org/abs/0912.3519}{{\tt 0912.3519}}.

\bibitem{Bena:2011wh}
  I.~Bena, G.~Giecold, M.~Grana, N.~Halmagyi, S.~Massai,
  ``The backreaction of anti-D3 branes on the Klebanov-Strassler geometry,''
 \href{http://arXiv.org/abs/1106.6165}{{\tt 1106.6165}}.

\bibitem{Dymarsky:2011pm}
 ÊA.~Dymarsky,
 Ê``On gravity dual of a metastable vacuum in Klebanov-Strassler theory,''
\href{http://arxiv.org/abs/1102.1734}{{\tt 1102.1734}}.
%%CITATION = ARXIV:1102.1734;%%


\bibitem{Klebanov:2010qs}
I.~R. Klebanov and S.~S. Pufu, ``{M-Branes and Metastable States},''
\href{http://arXiv.org/abs/1006.3587}{{\tt 1006.3587}} [hep-th].
%%CITATION = 1006.3587;%%.

\bibitem{Stenzel:1993}
M.~Stenzel, ``Ricci-flat metrics on the complexification of a compact rank one
  symmetric space,'' {\em Manuscripta Math.} {\bf 80} 1 (1993).

\bibitem{Cvetic:2000db}
M.~Cvetic, G.~W. Gibbons, H.~Lu and C.~N. Pope, ``{Ricci-flat metrics,
  harmonic forms and brane resolutions},'' {\em Commun. Math. Phys.} {\bf 232}
  (2003) 457--500,
\href{http://arXiv.org/abs/hep-th/0012011}{{\tt hep-th/0012011}}.
%%CITATION = HEP-TH/0012011;%%.

\bibitem{Bena:2010gs}
I.~Bena, G.~Giecold, and N.~Halmagyi, ``{The Backreaction of Anti-M2 Branes on
  a Warped Stenzel Space},''
\href{http://arXiv.org/abs/1011.2195}{{\tt 1011.2195}}.
%%CITATION = 1011.2195;%%.

\bibitem{Borokhov:2002fm}
V.~Borokhov and S.~S. Gubser, ``{Non-Supersymmetric Deformations of the Dual of
  a Confining Gauge Theory},'' {\em JHEP} {\bf 05} (2003) 034,
\href{http://arXiv.org/abs/hep-th/0206098}{{\tt hep-th/0206098}}.
%%CITATION = HEP-TH/0206098;%%.

\bibitem{Giecold:toappear}
G.~Giecold and N. Halmagyi, unpublished.

%\cite{Giecold:2011gw}
\bibitem{Giecold:2011gw}
  G.~Giecold, E.~Goi and F.~Orsi,
  ``Assessing a candidate IIA dual to metastable supersymmetry-breaking,''
 \href{http://arXiv.org/abs/1108.1789}{{\tt 1108.1789}}.

%\cite{Martelli:2009ga}
\bibitem{Martelli:2009ga}
  D.~Martelli and J.~Sparks,
  ``AdS(4) / CFT(3) duals from M2-branes at hypersurface singularities and
  their deformations,''
  JHEP {\bf 0912} (2009) 017
 \href{http://arXiv.org/abs/0909.2036}{{\tt 0909.2036}}.


%\cite{Bena:2010ze}
\bibitem{Bena:2010ze}
  I.~Bena, G.~Giecold, M.~Grana, N.~Halmagyi,
  ``On The Inflaton Potential From Antibranes in Warped Throats,''
  \href{http://arXiv.org/abs/1011.2626}{{\tt 1011.2626}}.


\bibitem{Bena:2011hz}
  I.~Bena, G.~Giecold, M.~Grana, N.~Halmagyi, S.~Massai,
  ``On Metastable Vacua and the Warped Deformed Conifold: Analytic
  Results,''
\href{http://arXiv.org/abs/1102.2403}{{\tt 1102.2403}}.

%\cite{Gukov:1999ya}
\bibitem{Gukov:1999ya}
  S.~Gukov, C.~Vafa and E.~Witten,
  ``CFT's from Calabi-Yau four folds,''
  Nucl.\ Phys.\  B {\bf 584} (2000) 69
  [Erratum-ibid.\  B {\bf 608} (2001) 477]
 \href{http://arXiv.org/abs/hep-th/9906070}{{\tt hep-th/9906070}}.

%\cite{Aharony:2009fc}
\bibitem{Aharony:2009fc}
  O.~Aharony, A.~Hashimoto, S.~Hirano and P.~Ouyang,
  ``D-brane Charges in Gravitational Duals of 2+1 Dimensional Gauge Theories
  and Duality Cascades,''
  JHEP {\bf 1001} (2010) 072
 \href{http://arXiv.org/abs/0906.2390}{{\tt 0906.2390}}.

%\cite{Hashimoto:2011aj}
\bibitem{Hashimoto:2011aj}
  A.~Hashimoto and P.~Ouyang,
  ``Quantization of charges and fluxes in warped Stenzel geometry,''
  JHEP {\bf 1106} (2011) 124
 \href{http://arXiv.org/abs/1104.3517}{{\tt 1104.3517}}.


\bibitem{Bergman:2001qi}
  A.~Bergman and C.~P.~Herzog,
  ``The Volume of some nonspherical horizons and the AdS / CFT
  correspondence,''
  JHEP {\bf 0201} (2002) 030 ,
 \href{http://arXiv.org/abs/hep-th/0108020}{{\tt hep-th/0108020}}.


%\cite{Hashimoto:2011nn}
\bibitem{Hashimoto:2011nn}
  A.~Hashimoto,
  ``Comments on domain walls in holographic duals of mass deformed conformal
  field theories,''
  JHEP {\bf 1107} (2011) 031
  [arXiv:1105.3687 [hep-th]].
  %%CITATION = JHEPA,1107,031;%%
 \href{http://arXiv.org/abs/1105.3687}{{\tt 1105.3687}}.

%\cite{McGuirk:2009xx}
\bibitem{McGuirk:2009xx}
  P.~McGuirk, G.~Shiu, Y.~Sumitomo,
  ``Non-supersymmetric infrared perturbations to the warped deformed conifold,''
  Nucl.\ Phys.\  {\bf B842 } (2010)  383-413.
\href{http://arXiv.org/abs/0910.4581}{{\tt 0910.4581}}.

%\cite{Blaback:2010sj}
\bibitem{Blaback:2010sj}
  J.~Blaback, U.~H.~Danielsson, D.~Junghans, T.~Van Riet, T.~Wrase, M.~Zagermann,
  ``Smeared versus localised sources in flux compactifications,''
  JHEP {\bf 1012 } (2010)  043.
\href{http://arXiv.org/abs/1009.1877}{{\tt 1009.1877}}.

%\cite{Blaback:2011nz}
\bibitem{Blaback:2011nz}
  J.~Blaback, U.~H.~Danielsson, D.~Junghans, T.~Van Riet, T.~Wrase, M.~Zagermann,
  ``The problematic backreaction of SUSY-breaking branes,''
   \href{http://arXiv.org/abs/1105.4879}{{\tt 1105.4879}}.

%\cite{Kachru:2003sx}
\bibitem{Kachru:2003sx}
  S.~Kachru, R.~Kallosh, A.~D.~Linde, J.~M.~Maldacena, L.~P.~McAllister and S.~P.~Trivedi,
  ``Towards inflation in string theory,''
  JCAP {\bf 0310} (2003) 013
 \href{http://arXiv.org/abs/hep-th/0308055}{{\tt hep-th/0308055}}.

%\cite{Pufu:2010ie}
\bibitem{Pufu:2010ie}
  S.~S.~Pufu, I.~R.~Klebanov, T.~Klose and J.~Lin,
  ``Green's Functions and Non-Singlet Glueballs on Deformed Conifolds,''
  J.\ Phys.\ A  {\bf 44} (2011) 055404
 \href{http://arXiv.org/abs/1009.2763}{{\tt 1009.2763}}.

%\cite{Grana:2000jj}
\bibitem{Grana:2000jj}
  M.~Grana and J.~Polchinski,
  ``Supersymmetric three form flux perturbations on AdS(5),''
  Phys.\ Rev.\  D {\bf 63} (2001) 026001
  \href{http://arXiv.org/hep-th/0009211}{{\tt hep-th/0009211}}.


 \end{thebibliography}
\end{document}